\documentclass[a4paper,12pt]{article}
\usepackage{woojung}
\title{A Bayesian Non-parametric Approach for Causal Mediation with a Post-treatment Confounder}

\author{
    Woojung Bae \\
    Department of Statistics \\
    University of Florida \\
    Gainesville, FL 32611 \\
    \texttt{woojung.bae@ufl.edu}
    \And
    Michael J. Daniels \\
    Department of Statistics \\
    University of Florida \\
    Gainesville, FL 32611 \\
    \texttt{daniels@ufl.edu}
    \And
    Michael G. Perri \\
    Department of Clinical and \\
    Health Psychology \\
    University of Florida \\
    Gainesville, FL 32611 \\
    \texttt{mperri@ufl.edu}
}
\makeatletter

\begin{document}

\maketitle

\setlength{\parskip}{2pt}
\setlength{\abovedisplayskip}{2pt}
\setlength{\belowdisplayskip}{2pt}
\setlength{\abovedisplayshortskip}{2pt}
\setlength{\belowdisplayshortskip}{2pt}

\begin{abstract}
    {
    \linespread{1}\selectfont
    We propose a new Bayesian non-parametric (BNP) method for estimating the causal effects of mediation in the presence of a post-treatment confounder. We specify an enriched Dirichlet process mixture (EDPM) to model the joint distribution of the observed data (outcome, mediator, post-treatment confounders, treatment, and baseline confounders). The proposed BNP model allows more confounder-based clusters than clusters for the outcome and mediator. For identifiability, we use the extended version of the standard sequential ignorability as introduced in \citet{hong2022posttreatment}. The observed data model and causal identification assumptions enable us to estimate and identify the causal effects of mediation, $i.e.$, the natural direct effects (NDE), and indirect effects (NIE). We conduct simulation studies to assess the performance of our proposed method. Furthermore, we apply this approach to evaluate the causal mediation effect in the Rural LITE trial, demonstrating its practical utility in real-world scenarios.
    }
\end{abstract}
    
\newpage
\section{Introduction} \label{EDPM.sec.1}
    
    Researchers in social, behavioral, and health sciences, are often interested in with making inferences about causal effects \citep{wu2011linking, woo2015distinct, zhao2018functional, sohn2019compositional}. Some of these are interested in inference on the casual treatment impact on an outcome mediated by intermediate variables \citep{kraemer2002mediators}. For decades, the process through which treatments affect outcomes through intermediate variables was a "black box" \citep{imai2011unpacking}, and various approaches have been proposed by researchers to examine it. 
    
    There is a vast frequentest literature on mediation analysis \citep{pearl2001direct, mackinnon2002comparison, robins2003semantics, preacher2004spss, petersen2006estimation, vanderweele2009marginal, imai2010identification, albert2011generalized, tchetgen2012semiparametric, valeri2013mediation}. Bayesian methods include \citet{yuan2009bayesian, elliott2010bayesian, schwartz2011bayesian, daniels2012bayesian, mattei2013exploiting, kim2017framework, kim2019bayesian}. In the context of the traditional structural equation model framework proposed by \citet{baron1986moderator}, \citet{yuan2009bayesian} employ Bayesian methods to calculate the posterior distribution of the product of coefficients, which allows for inference on the desired causal effects. However, the methods in \citet{elliott2010bayesian, schwartz2011bayesian, mattei2013exploiting} primarily aim to estimate a principal causal effect, which differs from our causal estimands of interest, namely the NDE and NIE. Moreover, Bayesian non-parametric (BNP) models have not been widely employed to estimate causal mediation effects. \citet{kim2017framework} established a BNP framework for assessing causal mediation effects. Specifically, they use a Dirichlet process mixture of multivariate normals (DPMN), as a prior for the joint distribution of the observed data (outcome, mediator, and confounders) to estimate causal mediation effects. They consider two sets of uncheckable assumptions, introduced in \citet{imai2010identification} and \citet{daniels2012bayesian}, respectively. The DPMN approach in \citet{kim2017framework} is not well-suited for situations with more than a few confounders, or non-continuous outcomes, mediators or confounders. Furthermore, the aforementioned assumptions are not valid when a post-treatment confounder is present.
    
    In the presence of post-treatment confounders, the standard cross-world counterfactual independence between the mediator and the outcome is often unrealistic \citep{hong2022posttreatment, rudolph2023efficient}. Consequently, the most common alternatives invoke an additional assumption regarding the post-treatment confounder, which treats it as the first mediator among multiple mediators in a sequential order. These assumptions are typically accompanied by a set of model-based assumptions within the generalized linear structural model framework \citep{albert2011generalized, albert2019generalized, daniel2015causal, imai2013identification}. Strategies for adjusting post-treatment confounding using weighting-based methods have also been developed \citep{huber2014identifying, hong2015causality, hong2018weighting}, which also invoke strong assumptions regarding the post-treatment confounder. Alternatively, many researchers propose different causal estimands that do not partition the average treatment effect (ATE) into natural indirect effect (NIE) and natural direct effect (NDE) \citep{geneletti2007identifying, didelez2012direct, vanderweele2014effect, lin2017interventional, miles2017quantifying, vansteelandt2017interventional, rudolph2018mediation, rudolph2018robust, miles2020semiparametric, wodtke2020effect, hejazi2020nonparametric, diaz2021nonparametric, nguyen2021clarifying, miles2022causal, tai2022identification}. Some of these studies have applied the interventional approach to derive new estimands for estimating direct and indirect effects without assuming the absence of post-treatment confounding factors. Specifically, recent work by \citet{hejazi2020nonparametric, diaz2021nonparametric} provide theoretical properties of the interventional effects within the nonparametric structural equation model (NPSEM) framework, along with assumptions that do not depend on cross-world counterfactual independence. 
    
    Recently, \citet{hong2022posttreatment} and \citet{rudolph2023efficient} propose assumptions for estimating NIE and NDE in the presence of post-treatment confounders. The proposed method in \citet{rudolph2023efficient} is designed to estimate NDE and NIE while taking into account intermediate confounding and monotonicity constraints when the post-treatment confounder is binary. However, when the post-treatment confounder is continuous, their assumptions are not sufficient. \citet{hong2022posttreatment} proposed a relaxed version of the cross-world counterfactual independence assumption, in conjunction with the ratio of mediator probability weighting (RMPW), which allows for treatment-by-mediator interaction \citep{hong2010ratio, hong2015causality, hong2012weighting, hong2015ratio, lange2012simple, tchetgen2012semiparametric}. Utilizing a Gaussian copula model \citep{nelsen2007introduction}, \citet{hong2022posttreatment} considers the conditional correlation between the post-treatment confounder based on the actual treatment condition and the counterfactual treatment condition, and provide a sensitivity analysis technique for addressing post-treatment confounding. 
    
    Here, we propose a BNP method for estimating causal effects of mediation in the presence of a post-treatment confounder. We specify an enriched Dirichlet process \citep[EDPM;][]{wade2011enriched, wade2014improving, roy2018bayesian} to model the joint distribution of the observed data, which includes the outcome, mediator, treatment, post-treatment confounder, and baseline confounders. The proposed BNP model allows for more confounder-clusters than clusters for the outcome and mediator. For identifiability, we consider an extended version of sequential ignorability for a post-treatment confounder \citep{hong2022posttreatment}. The combination of the observed data model and causal assumptions enables the estimation and identification of causal effects of mediation, $i.e.$, NDE and NIE. We employ a Gaussian copula model to model a post-treatment confounder, similar to the approach taken in previous studies \citep{daniels2012bayesian, kim2017framework, hong2022posttreatment}. This is because the post-treatment confounder is only partially observed, which means that if a subject is assigned to the treatment group, their potential post-treatment confounder value for the counterfactual control condition remains unobserved. A Gaussian copula model is useful for handling this type of partially `missing' data. While \citet{hong2022posttreatment} utilizes a parametric approach based on the ratio of mediator probability weighting (RMPW), our method enables more flexible estimation of the causal mediation effect and avoids the need for RMPW by sampling from the posterior distribution based on the EDPM. In addition, our method unlike \citet{hong2022posttreatment} allows for easy computation of NDE and NIE conditional on a subset of the confounding variables and any missingness in the observed data can be addressed under an assumption of ignorable missingness since we specify the joint distribution. 
    
    We applied the proposed method to data from the Rural LITE trial \citep{perri2014comparative}. The Rural LITE trial was conducted to examine the effects and costs of different behavioral weight loss and maintenance treatments delivered via Cooperative Extension Offices in rural communities. Subjects were randomized to either a low, moderate, or high dose of behavioral treatment or a control condition. Our focus is on the attendance rate as a potential mediator that can affect weight change by the end of the study. Weight was measured at months 0 (baseline), 6, and 24, and the primary outcome was the change in body weight from baseline to the final follow-up (month 24). The weight change from baseline to month 6 is a potential post-treatment confounder. We collected the following baseline confounders: sex, age, race, and BMI at baseline. 
    
    The remainder of the article is organized as follows. Section \ref{EDPM.sec.2} introduces the causal effects of interest and proposes a Bayesian nonparametric (BNP) model for the observed data. In Section \ref{EDPM.sec.3}, a set of assumptions is presented which are sufficient for identification. Posterior computation of the proposed BNP is discussed in Section \ref{EDPM.sec.4}. Section \ref{EDPM.sec.5} describes simulation studies that were conducted to evaluate the performance of the BNP model. The proposed approach is applied to estimate causal mediation effects in the Rural LITE data in Section \ref{EDPM.sec.6}. Finally, a discussion is provided in Section \ref{EDPM.sec.7}.
    
\section{Definition of Causal Effects and Specification of the Observed Data Model} \label{EDPM.sec.2}
    
    Let $Z_{i}$ denote a binary treatment indicator for the $i^{th}$ subject ($i = 1, \cdots , n$). Denote by $\boldsymbol{C}_{i}$ a $p_{\boldsymbol{C}} \times 1$ set of baseline confounders and let $V_{i}$ be a post-treatment confounder. Let $M_{i}$ denote the value of the mediator and let $Y_{i}$ be the value of the outcome. To define the causal mediation effects, we use the potential outcome framework \citet{rubin1974estimating}. Under the assumption of Stable Unit Treatment Value Assumption (SUTVA), \citep[]{rubin1980randomization}, potential values of the post-treatment confounder do not vary with treatments assigned to others. That is, for a vector of randomized treatments $\boldsymbol{Z}^{\left( n \right)} \in \left\{ 0,1 \right\}^{\otimes n}$, $V_{i, \boldsymbol{z}^{\left( n \right)}} = V_{i, Z_{i} }$ where $V_{i, \boldsymbol{z}^{\left( n \right)}}$ denotes the value of the post-treatment confounder $V_{i}$ for the $i^{th}$ subject that would have been observed had a vector of treatments $\boldsymbol{Z}^{\left( n \right)}$ been set to $\boldsymbol{z}^{\left( n \right)}$. Similarly, the random variable $M_{i, \boldsymbol{z}^{\left( n \right)}, \boldsymbol{v}^{\left( n \right)}}$ is equal to $M_{i, Z_{i}, V_{i}}$, where $M_{i, \boldsymbol{z}^{\left( n \right)}, \boldsymbol{v}^{\left( n \right)}}$ is the value of the mediator for the $i^{th}$ subject that would have been observed had a vector of treatments $\boldsymbol{Z}^{\left( n \right)}$ been set to $\boldsymbol{z}^{\left( n \right)}$ and values of the post-treatment confounder $\boldsymbol{V}^{\left( n \right)}$ been set to $\boldsymbol{v}^{\left( n \right)}$. Finally, the random variable $Y_{i, \boldsymbol{z}^{\left( n \right)}, \boldsymbol{v}^{\left( n \right)}, \boldsymbol{m}^{\left( n \right)}}$ is equal to $Y_{i, Z_{i}, V_{i}, M_{i}}$, where $Y_{i, \boldsymbol{z}^{\left( n \right)}, \boldsymbol{v}^{\left( n \right)}, \boldsymbol{m}^{\left( n \right)}}$ is the value of the outcome for the $i^{th}$ subject that would have been observed had a vector of treatments $\boldsymbol{Z}^{\left( n \right)}$ been set to $\boldsymbol{z}^{\left( n \right)}$, a vector of post-treatment confounders $\boldsymbol{V}^{\left( n \right)}$ been set to $\boldsymbol{v}^{\left( n \right)}$, and values of the mediator $\boldsymbol{M}^{\left( n \right)}$ been set to $\boldsymbol{m}^{\left( n \right)}$. To simplify notation and focus on evaluating the natural effects, the subject index $i$ is omitted throughout the paper, following the work of \citep{robins1992identifiability, pearl2001direct}. Furthermore, instead of using $M_{z, V_{z}}$ and $Y_{z, V_{z}, M_{z', V_{z'}}}$ for $z, z' \in \left\{ 0, 1 \right\}$, we write $M_{z}$ and $Y_{z, M_{z'}}$. The observed post-treatment confounder, mediator, and outcome are defined as follows: $V = Z V_{1} + \left( 1 - Z \right) V_{0}$, $M = Z M_{1} + \left( 1 - Z \right) M_{0}$, and $Y = Z Y_{1, M_{1}} + \left( 1 - Z \right) Y_{0, M_{0}}$. We define NDE and NIE conditional on baseline confounders $\boldsymbol{C} = \boldsymbol{c}$ as $\nnie \left( \boldsymbol{c} \right) = \nE \left[ Y_{1, M_{1}} - Y_{1, M_{0}} | \boldsymbol{c} \right] $ and $\nnde \left( \boldsymbol{c} \right) = \nE \left[ Y_{1, M_{0}} - Y_{0, M_{0}} | \boldsymbol{c} \right]$. The natural indirect effect, $\nnie \left( \boldsymbol{c} \right)$, quantifies the effect of the treatment through the mediator for a fixed value of confounders $\boldsymbol{C} = \boldsymbol{c}$. The natural direct effect, $\nnde \left( \boldsymbol{c} \right)$, quantifies the effect of the treatment on the outcome by setting the mediator $M$ to its natural value $M_{0}$ (the value of the mediator in the absence of the treatment) given a fixed value of the baseline confounders $\boldsymbol{C}$. The total effect is the sum of the two effects $\nate \left( \boldsymbol{c} \right) = \nnie \left( \boldsymbol{c} \right) + \nnde \left( \boldsymbol{c} \right) = \nE \left[ Y_{1, M_{1}} - Y_{0, M_{0}} | \boldsymbol{c} \right]$. After integrating out the baseline confounders, we obtain the marginal causal effects $\nnie$, NDE, and ATE. Our goal is to identify causal effects from the observed data $\left( Y, M, V, Z, \boldsymbol{C} \right)$.
    
    In what follows, we first provide details on BNP models for the observed data in Section \ref{EDPM.subsec.2.1}. 
    
\subsection{A Bayesian non-parametric model for the observed data} \label{EDPM.subsec.2.1}
    
    To estimate the causal effects introduced in Section \ref{EDPM.sec.3}, we first need to estimate the joint distribution $\nP \left( Y, M, V, Z, \boldsymbol{C} \right)$. Let $\boldsymbol{X} = \left( Z, \boldsymbol{C}^{\top} \right)^{\top}$, and consider $\nP \left( Y, M, V, \boldsymbol{X} \right)$. We propose modeling the joint distribution of $\left( Y, M, V, \boldsymbol{X} \right)$ using an EDPM \citep{wade2011enriched, wade2014improving, roy2018bayesian} specified as follows,
    \begin{align}
        Y_{i} | M_{i}, V_{i}, \boldsymbol{X}_{i}; \theta_{i}^{y} & \sim f \left( y_{i} | m_{i}, v_{i}, \boldsymbol{x}_{i}; \theta_{i}^{y} \right) \nonumber \\
        M_{i} | V_{i}, \boldsymbol{X}_{i}; \theta_{i}^{m} & \sim f \left( m_{i} | v_{i}, \boldsymbol{x}_{i}; \theta_{i}^{m} \right) \nonumber \\
        V_{i} | \boldsymbol{X}_{i}; \omega_{i}^{v} & \sim f \left( v_{i} | \boldsymbol{x}_{i}; \omega_{i}^{v} \right) \nonumber \\
        \boldsymbol{X}_{i,q}; \omega_{i,q}^{x} & \sim f \left( x_{i,q};  \omega_{i,q}^{x} \right), \;\; q=1, \cdots , p_{1}+p_{2}+1  \nonumber \\
        \left( \theta_{i} , \omega_{i} \right) | \nG & \sim \nG \nonumber \\
        \nG & \sim \nedp \left( \alpha^{\theta}, \alpha^{\omega}, \nG_{0} \right) \label{EDPM.eq.2.0.1}
    \end{align}
    where $\theta_{i} = \left( \theta_{i}^{y}, \theta_{i}^{m} \right)$ and $\omega_{i} = \left( \omega_{i}^{v}, \omega_{i}^{x} \right)$. The notation $\nG \sim \nedp \left( \alpha^{\theta}, \alpha^{\omega}, \nG_{0} \right)$ means that $\nG^{\theta} \sim \ndp \left( \alpha^{\theta}, \nG_{0}^{\theta} \right)$ and $\nG^{\omega|\theta} \sim \ndp \left( \alpha^{\theta}, \nG_{0}^{\omega | \theta} \right)$ with base measure $\nG_{0} = \nG_{0}^{\theta} \times \nG_{0}^{\omega | \theta}$. 
    
    In the EDPM specification, each subject $i$ has its own parameter $\left( \theta_{i}, \omega_{i} \right)$, but subjects in the same cluster share the same parameter values, $\left( \theta_{i}, \omega_{i} \right)$, due to the discreteness of $\nG$ \citep{ferguson1973bayesian}. Baseline covariates, $\boldsymbol{X}$, are assumed to be independent within clusters, while local dependence is allowed for the post-treatment confounder $V$. The EDPM has two concentration parameters $\alpha^{\theta}$ and $\alpha^{\omega}$, while Dirichlet Process mixture (DPM) models only have one concentration parameter. The number of $y \& m$-clusters depends on the concentration parameter $\alpha^{\theta}$, and the number of nested clusters, $v \& \boldsymbol{x}$-clusters given $y \& m$-clusters, depends on the concentration parameter $\alpha^{\omega}$. Lower values of $\alpha^{\theta}$ and $\alpha^{\omega}$ indicate fewer clusters, similar to the concentration parameter in DPM. This allows for more $v \& \boldsymbol{x}$-clusters than $y \& m$-clusters, which is crucial because the dimension of $v \& \boldsymbol{x}$ is typically (much) larger than that of $y \& m$. For the sake of simplicity, $y$-clusters and $\boldsymbol{x}$-clusters will be used instead of $y \& m$-clusters and $v \& \boldsymbol{x}$-clusters, respectively, in the following sections.
    
    Within each of the $y$-clusters, we assume two generalized linear models, one for $f \left( y | m, v, \boldsymbol{x}; \theta^{y} \right)$ and for $f \left( m | v, \boldsymbol{x}; \theta^{y} \right)$. If $Y$ is continuous, we specify $Y | \mathbb{M}; \beta^{y}, \sigma^{y,2} \sim \mnormals{\mathbb{M} \beta^{y}, \sigma^{y,2}} $ where $\theta^{y} = \left( \beta^{y}, \sigma^{y,2} \right)$ and $\mathbb{M} = \left( 1, M, V, Z, \boldsymbol{C}^{\top} \right)^{\top}$ (a design matrix involving $M$, $V$ and $\boldsymbol{X}$). One may specify $Y | \mathbb{M}; \theta^{y} \sim \mbernoulli{\minvprobit{\mathbb{M} \theta^{y}}} $ if $Y$ is binary. Similarly, we assume a GLM for $f \left( m | v, \boldsymbol{x}; \theta^{m} \right)$ within each of the $y$-clusters ($y \& m$-clusters). For a post-treatment confounder $V$, we specified a local GLM given baseline confounders $\boldsymbol{X}$ and treatment $Z$. Similar to \citep{roy2018bayesian}, baseline confounders $\boldsymbol{X}$ are assumed to be locally independent. It is worth noting that all variables are globally dependent and may have nonlinear relationships even though we assume a parametric model for each $y$, $m$, $v$, and $\boldsymbol{x}$ within each cluster.
    
    The EDPM model \eqref{EDPM.eq.2.0.1} has a square-breaking representation \citep{wade2011enriched},
    \begin{gather*}
        \nP \left( y, m, v, \boldsymbol{x} | \nG \right) = \sum_{k=1}^{\infty} \gamma_{k} f \left( y | m, v, \boldsymbol{x}; \theta_{k}^{y} \right) f \left( m | \boldsymbol{x}; \theta_{k}^{m} \right) \sum_{r=1}^{\infty} \gamma_{r | k} f \left( v | \boldsymbol{x}; \omega_{r|k}^{v} \right) f \left( \boldsymbol{x}; \omega_{r|k}^{x} \right), 
    \end{gather*}
    where $k$ indexes the $y$-clusters and the $f \left( \cdot \right)$ are the corresponding distribution. The weights have priors $\gamma_{k}' \sim \mbeta{1, \alpha^{\theta}} $ and $\gamma_{r | k}' \sim \mbeta{1, \alpha^{\omega}} $ where $\gamma_{k} = \gamma_{k}' \prod_{l < k } \left( 1 - \gamma_{l}' \right)$ and $\gamma_{r | k} = \gamma_{r | k}' \prod_{d < r } \left( 1 - \gamma_{d | k}' \right)$. 
    
    From the joint distribution, we can derive the following forms for the conditional distributions (which we use in the G-computation algorithm in Section \ref{EDPM.sec.4.1}),
    \begin{gather*}
        \nP \left( y | m, v, \boldsymbol{x} \right) = \sum_{k=1}^{ \infty} \Lambda_{k}^{y} (m, v, \boldsymbol{x}) f \left( y | m, v, \boldsymbol{x}; \theta_{k}^{y} \right), 
    \end{gather*}
    where
    \begin{gather*}
        \Lambda_{k}^{y} \left( m, v, \boldsymbol{x} \right) = \frac{\gamma_{k} f \left( m | \boldsymbol{x}; \theta_{k}^{m} \right) \sum_{r=1}^{\infty} \gamma_{r | k} f \left( v | \boldsymbol{x}; \omega_{k}^{V} \right) f \left( \boldsymbol{x}; \omega_{r|k}^{x} \right)}{\sum_{d=1}^{\infty} \gamma_{d} f \left( m | \boldsymbol{x}; \theta{d}^{m} \right) \sum_{r=1}^{\infty} \gamma_{r | d} f \left( v | \boldsymbol{x}; \omega_{k}^{V} \right) f \left( \boldsymbol{x}; \omega_{r|d} \right)} 
        = \frac{\lambda_{k}^{y} \left( m, v, \boldsymbol{x} \right)}{\sum_{d=1}^{\infty} \lambda_{d}^{y} \left( m, v, \boldsymbol{x} \right)}, 
    \end{gather*}
    The weights $\Lambda_{k}^{y} \left( m, v, \boldsymbol{x} \right)$ depends on $m, v, \boldsymbol{x}$. Note that $ \nP \left( m | v, \boldsymbol{x} \right) $, and $ \nP \left( v | \boldsymbol{x} \right) $ have similar forms; see the supplementary materials for details. Hence, $ \nP \left( y | m, v, \boldsymbol{x} \right) $, $ \nP \left( m | v, \boldsymbol{x} \right) $, and $ \nP \left( v | \boldsymbol{x} \right) $ are computationally tractable, flexible, non-linear, non-additive models even though locally each of $f \left( y | m, v, \boldsymbol{x}; \theta^{y} \right)$, $f \left( m | v, \boldsymbol{x}; \theta^{m} \right)$, and $f \left( v | \boldsymbol{x}; \omega^{v} \right)$ is assumed to follow a simple GLM.

\section{Identifying Assumptions and Inference on Causal Effects} \label{EDPM.sec.3}
    
    In this section, we define the causal effects of interest and the identification assumptions. It is important to note that estimating the joint distribution of the observed data is a separate step from defining the causal effects and specifying the identifying assumptions. We employ a set of assumptions introduced in \citet{hong2022posttreatment}, which are an extension of the standard sequential ignorability in \citep{imai2010identification}, to assess the causal effects of interest given the observed data distribution with a post-treatment confounder. 
    
    The causal graph \ref{EDPM.ce.post} below shows the causal mediation structure and with a post-treatment confounder, $V$ and a randomized treatment, $Z$.

    \citet{hong2022posttreatment} modify the sequential ignorability assumptions of \citet{imai2010identification} as follows, for $z, z'= 0, 1$ and $z \ne z '$: 
    \begin{enumerate}[leftmargin=1.5em]
        \item[\eqref{EDPM.eq.3.1}] \textit{Ignorable treatment assignment given the observed baseline confounders:} 
        \begin{align}  \label{EDPM.eq.3.1}
            \left\{ Y_{z, m}, M_{z}, M_{z'}, V_{z}, V_{z'} \right\} \indep Z | \boldsymbol{C} = \boldsymbol{c}. \tag{1}
        \end{align}
        Given that $V_{z}$ and $V_{z'}$ are each specified under a fixed treatment condition, treatment assignment is also ignorable given the potential post-treatment confounder values $V_{z} = v$ and $V_{z'} = v'$,
        \begin{align} \label{EDPM.eq.3.1.2}
            \left\{ Y_{z, m}, M_{z}, M_{z'} \right\} \indep Z |  V_{z}, V_{z'}, \boldsymbol{C} = \boldsymbol{c}. \tag{1'}
        \end{align}
        Assumptions \eqref{EDPM.eq.3.1} and \eqref{EDPM.eq.3.1.2} are guaranteed when the treatment is randomized. Assumption \eqref{EDPM.eq.3.1} could be too strong in a quasi-experimental study in which treatment selection might still be associated with unobserved baseline confounders within levels of $\boldsymbol{C}$.
        
        \item[\eqref{EDPM.eq.3.2}] \textit{Ignorable mediator value assignment under each treatment condition given the observed baseline confounders and a post-treatment confounder:} 
        \begin{align} \label{EDPM.eq.3.2}
            Y_{z, m}  \indep \left\{ M_{z}, M_{z'} \right\} | V_{z} = v, Z = z, \boldsymbol{C} = \boldsymbol{c}. \tag{2}
        \end{align}
        Assumption \eqref{EDPM.eq.3.2} precludes cross-world connections between the mediator and the outcome. This assumption is more reasonable than the standard ignorability assumption \eqref{EDPM.eq.3.1.2}, which only conditions on baseline confounders, because post-treatment confounding is often unavoidable. Under assumption \eqref{EDPM.eq.3.2}, subjects with identical baseline confounder values $\boldsymbol{C} = \boldsymbol{c}$ and identical post-treatment $V_{z} = v$ are assumed to have independent potential outcomes $Y_{z, m}$ for each of the two treatment conditions $M_{z}$ and $M_{z'}$. However, clearly if the mediator-outcome relationship is confounded by an unmeasured baseline or post-treatment confounder, assumption \eqref{EDPM.eq.3.2} may not hold.
        
        \item[\eqref{EDPM.eq.3.3}] \textit{Conditional cross-world independence between the post-treatment confounder and the mediator:} 
        \begin{align} \label{EDPM.eq.3.3}
            M_{z'} \indep V_{z} | V_{z'} = v', Z = z', \boldsymbol{C} = \boldsymbol{c}. \tag{3}
        \end{align}
        Assumption \eqref{EDPM.eq.3.3} implies that, for a given treatment condition $z'$ and known values of $V_{z'}$ and $\boldsymbol{C}$, $V_{z}$ does not contain any additional information about $M_{z'}$. This assumption may also not hold if there is an unmeasured baseline or post-treatment confounder that affects $M_{z'}$ and $V_{z}$ but not $V_{z'}$.
        
        \item[\eqref{EDPM.eq.3.4}] \textit{Gaussian Copula:} The joint distribution of two potential post-treatment confounders conditional on baseline confounders is assumed to follow a Gaussian copula model \citep{nelsen2007introduction}:  
        \begin{align} \label{EDPM.eq.3.4}
            F_{V_{z'}, V_{z}} \left( v_{z'}, v_{z} | \boldsymbol{C} = \boldsymbol{c} \right) = \Phi_{2} \left( \Phi_{1}^{-1} \left( F_{V_{z'}} \left( v_{z'} | \boldsymbol{C} = \boldsymbol{c} \right) \right), \Phi_{1}^{-1} \left( F_{V_{z}} \left( v_{z} | \boldsymbol{C} = \boldsymbol{c} \right) \right) \right), \tag{4}
        \end{align}
        where $\Phi_{1}$ is the univariate standard normal cumulative distribution function (CDF) and $\Phi_{2}$ is the bivariate normal CDF with mean $\boldsymbol{0}$, variance $\boldsymbol{1}$, and correlation $\rho \in \left( -1, 1 \right)$. Importantly, this formulation of the joint distribution for the potential post-treatment confounders imposes no constraints on the models, $V_{z'} | \boldsymbol{C}$ and $V_{z} | \boldsymbol{C}$, which we estimate using the EDPM.
        
        Since we never observe $V_{z'}$ and $V_{z}$ simultaneously, estimating $\rho$ from the data is not feasible. Therefore, one may treat $\rho$ as known and include it in a sensitivity analysis \citep{daniels2012bayesian}, or specify a prior, such as $\munif{0,1}$ if assuming only a positive correlation between the potential post-treatment confounders.
        
    \end{enumerate}
    
    \citet{hong2022posttreatment} prove the following under assumptions \eqref{EDPM.eq.3.1} and \eqref{EDPM.eq.3.3}. 
    \begin{theorem} \label{EDPM.thm.3.1}
        Under assumptions \eqref{EDPM.eq.3.1} and \eqref{EDPM.eq.3.3}, 
        \begin{align*}
            & \nP \left( M_{z'} = m' | V_{z} = v, Z = z, \boldsymbol{C} = \boldsymbol{c} \right) \\
            & =
            \int \nP \left( M_{z'} = m' | V_{z'} = v', Z = z', \boldsymbol{C} = \boldsymbol{c} \right) \nP \left( V_{z'} = v' | V_{z} = v, Z = z, \boldsymbol{C} = \boldsymbol{c} \right) dv'.
        \end{align*}
    \end{theorem}
    This theorem implies we can identify the conditional probability of a certain mediator value, $M_{z'} = m'$, under the counterfactual condition $Z = z$ via marginalizing out the conditional distribution, $V_{z'} | V_{z} = v, Z = z, \boldsymbol{C} = \boldsymbol{c}$. Note that the conditional distribution, $V_{z'} | V_{z} = v, Z = z, \boldsymbol{C} = \boldsymbol{c}$, is partially identified by \eqref{EDPM.eq.3.4}.
    
    The following theorem provides identification of the NIE and NDE. 
    \begin{theorem} \label{EDPM.thm.3.2}
        Under assumptions \eqref{EDPM.eq.3.1}, \eqref{EDPM.eq.3.2}, \eqref{EDPM.eq.3.3}, and \eqref{EDPM.eq.3.4}, NIE and NDE can be identified: for $z \ne z'$ and $v \ne v'$, 
        \begin{align*}
            \mE{ Y_{z, M_{z'}} | V_{z} = v, \boldsymbol{C} = \boldsymbol{c}}
            & =
            \iint \mE{ Y_{z, m'} | M_{z} = m', V_{z} = v, Z = z, \boldsymbol{C} = \boldsymbol{c} } \\
            & \hspace{5em} \ndF_{M_{z'} | V_{z'} = v', Z = z', \boldsymbol{C} = \boldsymbol{c}}  \left( m' \right) \ndF_{V_{z'} | V_{z} = v, Z = z, \boldsymbol{C} = \boldsymbol{c}}  \left( v' \right). 
        \end{align*}
        \begin{proof} 
            \begin{align*}
                & \mE{ Y_{z, M_{z'}} | V_{z} = v, \boldsymbol{C} = \boldsymbol{c}}  \\
                & =
                \int \mE{ Y_{z, m'} | M_{z'} = m' , V_{z} = v, \boldsymbol{C} = \boldsymbol{c} } \ndF_{M_{z'} | V_{z} = v, \boldsymbol{C} = \boldsymbol{c}} \left( m' \right)  \\
                & \overset{\eqref{EDPM.eq.3.1}}{=}
                \int \mE{ Y_{z, m'} | M_{z'} = m' , V_{z} = v, Z = z, \boldsymbol{C} = \boldsymbol{c} } \ndF_{M_{z'} | V_{z} = v, Z = z, \boldsymbol{C} = \boldsymbol{c}} \left( m' \right)  \\
                & \overset{\eqref{EDPM.eq.3.2}}{=}
                \int \mE{ Y_{z, m'} | M_{z} = m' , V_{z} = v, Z = z, \boldsymbol{C} = \boldsymbol{c} } \ndF_{M_{z'} | V_{z} = v, Z = z, \boldsymbol{C} = \boldsymbol{c}} \left( m' \right)   \\
                & \overset{\ref{EDPM.thm.3.1}}{=}
                \int \mE{ Y_{z, m'} | M_{z} = m' , V_{z} = v, Z = z, \boldsymbol{C} = \boldsymbol{c} } \int \ndF_{M_{z'} | V_{z'} = v', Z = z', \boldsymbol{C} = \boldsymbol{c}}  \left( m' \right) \ndF_{V_{z'} | V_{z} = v, Z = z, \boldsymbol{C} = \boldsymbol{c}} \left( v' \right)  \\
                & =
                \iint \mE{ Y_{z, m'} | M_{z} = m' , V_{z} = v, Z = z, \boldsymbol{C} = \boldsymbol{c} } \ndF_{M_{z'} | V_{z'} = v', Z = z', \boldsymbol{C} = \boldsymbol{c}}  \left( m' \right) \ndF_{V_{z'} | V_{z} = v, Z = z, \boldsymbol{C} = \boldsymbol{c}}  \left( v' \right)
            \end{align*}
        \end{proof}
    \end{theorem}

    The proof of Theorem \ref{EDPM.thm.3.2}, \citet{hong2022posttreatment} obtained the following results:
    \begin{align*}
        \mE{ Y_{z, M_{z'}} | V_{z} = v, \boldsymbol{C} = \boldsymbol{c}} 
        & =
        \iiint\psi \left( m', v, v', \boldsymbol{c}  \right) \mE{ Y_{z, m'} | M_{z} = m' , V_{z} = v, Z = z, \boldsymbol{C} = \boldsymbol{c} } \\
        & \hspace{8em} \ndF_{M_{z} | V_{z} = v, Z = z, \boldsymbol{C} = \boldsymbol{c}} \left( m' \right) \ndF_{V_{z'} | V_{z} = v, Z = z, \boldsymbol{C} = \boldsymbol{c}} \left( v' \right).
    \end{align*}
    where $\psi \left( m', v, v', \boldsymbol{c}  \right) = \frac{\nP \left( M_{z'} = m' | V_{z'} = v', Z = z', \boldsymbol{C} = \boldsymbol{c} \right)}{\nP \left( M_{z} = m' | V_{z} = v, Z = z, \boldsymbol{C} = \boldsymbol{c} \right)}$ by multiplying $\frac{\nP \left( M_{z} = m' | V_{z} = v, Z = z, \boldsymbol{C} = \boldsymbol{c} \right)}{\nP \left( M_{z} = m' | V_{z} = v, Z = z, \boldsymbol{C} = \boldsymbol{c} \right)} (= 1)$ to the integrand of the result of Theorem \ref{EDPM.thm.3.2}. They use a semiparametric approach with re-weighting. Since we have default flexible specifications for all the distribution in Theorem \ref{EDPM.thm.3.2}, this re-weighting step is unnecessary. Moreover, using our approach, it is easy to compute NDE and NIE conditional on (a subset of) confounders unlike \citet{hong2022posttreatment} and any missingness of $\left( Y, M, V, \boldsymbol{X} \right)$ where $\boldsymbol{X} = \left( Z, \boldsymbol{C}^{\top} \right)^{\top}$ can be handled via data augmentation under an ignorable missingness assumption.
    
\section{Posterior Computation} \label{EDPM.sec.4}
    We use the algorithm from \citet{wade2014improving}, which was an extension of \cite[Algorithm 8][]{neal2000markov} to generate posterior samples from the parameters in the observed data model; see the supplementary materials for details. Posterior distributions of the NIE, NDE, and ATE, as well as the corresponding point estimates and credible intervals, are computed via performing several post-processing steps for each posterior sample of the observed data model parameters. 
    
\subsection{Post-Processing Steps for Estimation of Causal Effects} \label{EDPM.sec.4.1}
    Any functionals of the distribution of potential outcomes such as $\nE \left[ Y_{z, M_{z'}} \right]$, $\nE \left[ Y_{z, M_{z'}} | V \right]$, or $\nE \left[ Y_{z, M_{z'}} | \boldsymbol{C} \right]$ can be computed using the parameters from the posterior distribution and the identifying assumptions in Section \ref{EDPM.sec.3}. The following shows how to compute the expectations of the potential outcomes, $\nE \left[ Y_{z, M_{z'}} \right]$,
    \begin{align*}
        \mE{ Y_{z, M_{z'}} | V_{z} = v, \boldsymbol{C} = \boldsymbol{c}}
        & =
        \iint \mE{ Y_{z, m'} | M_{z} = m', V_{z} = v, Z = z, \boldsymbol{C} = \boldsymbol{c} } \\
        & \hspace{5em} \ndF_{M_{z'} | V_{z'} = v', Z = z', \boldsymbol{C} = \boldsymbol{c}}  \left( m' \right) \ndF_{V_{z'} | V_{z} = v, Z = z, \boldsymbol{C} = \boldsymbol{c}}  \left( v' \right). 
    \end{align*}
    Let $\theta$ denote the parameters of outcome \& mediator regressions and $\omega$ denote the parameters for treatment and confounders (including post-treatment and baseline confounders). Also, let $K$ be the number of non-empty $y$-clusters ($y\&m$-clusters), and $K_{r|k}$ denote the number of non-empty $x$-clusters ($z\&x$-clusters) within the $k^{th}$ $y$-clusters. The form of the conditional expectation conditional on a posterior sample of $\left\{ \theta^{*}, \omega^{*}, s \right\}$ is
    \begin{align}
        & \mE{ Y | M = m', V = v, Z = z, \boldsymbol{C} = \boldsymbol{c}; \theta^{*}, \omega^{*}, s } \nonumber \\
        & = \frac{\lambda_{K + 1}^{y} \left( m', v, z, \boldsymbol{c} \right) \nE_{0} \left[ y | m', v, z, \boldsymbol{c} \right] + \sum_{l=1}^{K} \lambda_{l}^{y} \left( m', v, z, \boldsymbol{c} \right) \nE \left[ y | m', v, z, \boldsymbol{c}; \theta_{l}^{y*} \right]}{\lambda_{K + 1}^{y} \left( m', v, z, \boldsymbol{c} \right) + \sum_{l=1}^{K} \lambda_{l}^{y} \left( m', v, z, \boldsymbol{c} \right) } \label{EDPM.eq.4.1.1}
    \end{align}
    where $\lambda_{K + 1}^{y} \left( m', v, z, \boldsymbol{c} \right) = \frac{\alpha^{\theta}}{\alpha^{\theta} + n} f_{0} \left( m' | v, z, \boldsymbol{c} \right) f_{0} \left( v| z, \boldsymbol{c} \right) f_{0} \left( z, \boldsymbol{c} \right)$, and
    \begin{align*}
        \lambda_{l}^{y} \left( m', v, z, \boldsymbol{c} \right) = \frac{n_{l}}{\alpha^{\theta} + n} f \left( m' | v, z, \boldsymbol{c}; \theta_{l}^{m*} \right) \left\{ \frac{\alpha^{\omega}}{\alpha^{\omega} + n_{l}} f_{0} \left( v, z, \boldsymbol{c} \right) + \sum_{r=1}^{K_{l}} \frac{n_{r|l}}{\alpha^{\omega} + n_{l}} f \left( v, z, \boldsymbol{c}; \omega_{r|l}^{*} \right) \right\}.
    \end{align*}
    After marginalizing out the parameters over the prior distributions, we can get the distributions and mean, $f_{0} \left( z, \boldsymbol{c} \right)$, $f_{0} \left( v, z, \boldsymbol{c} \right)$, $f_{0} \left( m' | v, z, \boldsymbol{c} \right)$, and $\nE_{0} \left[ y | m', v, z, \boldsymbol{c} \right]$, respectively: $f_{0} \left( z, \boldsymbol{c} \right) = \int f \left( z, \boldsymbol{c}; \omega^{x} \right) d \nG_{0} \left( \omega^{x} \right)$, $f_{0} \left( v | z, \boldsymbol{c} \right) = \int f \left( v | z, \boldsymbol{c}; \omega \right) d \nG_{0} \left( \omega^{v} \right)$, $f_{0} \left( m' | v, z, \boldsymbol{c} \right) = \int f \left( m' | v, z, \boldsymbol{c}; \theta^{m} \right) d \nG_{0} \left( \theta^{m} \right)$, and $\nE_{0} \left[ y | m', v, z, \boldsymbol{c} \right] = \int \nE \left[ y | m', v, z, \boldsymbol{c}; \theta^{y} \right] d \nG_{0} \left( \theta^{y} \right)$. We apply MC integration to marginalize out $\left( V, \boldsymbol{C}, \boldsymbol{S} \right)$ given current values of the parameters, $\left\{ \theta^{*}, \omega^{*}, s \right\}$. Given the parameters from the Gibbs sampler, $\left\{ \theta^{*}, \omega^{*}, s \right\}$, we can sample $m$ from $\nP \left( m | V = v, Z = z, \boldsymbol{C} = \boldsymbol{c}; \theta^{*}, \omega^{*}, s \right)$; see the supplementary materials for details. We also can sample $v$ from $\nP \left( v | Z = z, \boldsymbol{C} = \boldsymbol{c}; \theta^{*}, \omega^{*}, s \right)$ defined as
    \begin{align} \label{EDPM.eq.4.1.2}
        & \nP \left( v | Z = z, \boldsymbol{C} = \boldsymbol{c}; \theta^{*}, \omega^{*}, s \right) \nonumber \\
        & = \frac{\lambda_{K_{K} + 1}^{v} \left( z, \boldsymbol{c} \right) f_{0} \left( v | z, \boldsymbol{c} \right) + \sum_{l=1}^{K} \sum_{r=1}^{K_{l}} \lambda_{r|l}^{v} \left( z, \boldsymbol{c} \right) f \left( v | z, \boldsymbol{c}; \omega_{r|l}^{v*} \right) }{ \lambda_{K_{K} + 1}^{v} \left( z, \boldsymbol{c} \right) + \sum_{l=1}^{K} \sum_{r=1}^{K_{l}} \lambda_{r|l}^{v} \left( z, \boldsymbol{c} \right)},
    \end{align}
    where $\lambda_{K_{K} + 1}^{v} \left( z, \boldsymbol{c} \right) = \left\{ \frac{\alpha^{\theta}}{\alpha^{\theta} + n} + \sum_{l=1}^{K} \frac{n_{l}}{\alpha^{\theta} + n} \frac{\alpha^{\omega}}{\alpha^{\omega} + n_{l}} \right\} f_{0} \left( z, \boldsymbol{c} \right)$, and $\lambda_{r|l}^{v} \left( z, \boldsymbol{c} \right) = \frac{n_{l}}{\alpha^{\theta} + n} \frac{n_{r|l}}{\alpha^{\omega} + n_{l}} f \left( z, \boldsymbol{c}; \omega_{r|l}^{x*} \right)$. To sample $v'$ from $F_{V' | V = v, Z = z, \boldsymbol{C} = \boldsymbol{c}; \rho} \left( v' \right)$, we use Gaussian Copula model in \eqref{EDPM.eq.3.4},
    \begin{gather} \label{EDPM.eq.4.1.3}
        F_{V' | V} \left( v' | v ; \rho \right) = \Phi \left( \frac{\Phi^{-1} \left( F_{V'} \left( v' \right) \right) - \rho \Phi^{-1} \left( F_{V} \left( v \right) \right)}{\sqrt{1 - \rho^{2}}} \right),
    \end{gather}
    where conditioning variables are omitted for simplicity. First, sample $u$ from a normal distribution with mean $\rho \Phi^{-1} \left( F_{V} \left( v \right) \right)$ and variance $1 - \rho^{2}$, and then set $v' = F_{V'}^{-1} \left( \Phi \left( u \right) \right)$. Note that $F_{V'}^{-1}$ is difficult to compute, but we can use standard optimization techniques, such as bisection or Newton-Raphson method, to compute $\hat{v}'$ such that $F_{V'} \left( \hat{v}' \right) = \Phi \left( u \right)$. Practically, we find an optimal value, $\hat{v}'$, such that $\norm{F_{V'} \left( \hat{v}' \right) - \Phi \left( u \right)}_{2} < \epsilon_{\text{tol}}$ where $\epsilon_{\text{tol}}$ is given tolerance. For a binary post-treatment confounder, one can follow the method in \citet{daniels2012bayesian} and \citet{kim2017framework}. 
    
    Here, we briefly describe the post-processing steps; see the supplementary materials for details. In particular, we draw $D$ samples using each posterior sample as follows: (1) draw $s_{\left( d \right)}^{y}$ and $s_{\left( d \right)}^{x}$ from a multinomial distribution (if new clusters are opened up, draw new parameters from the prior distribution), (2) draw $\boldsymbol{c}_{\left( d \right)}$ from $f \left( \boldsymbol{x}; \omega_{s_{\left( d \right)}^{\boldsymbol{x}} | s_{\left( d \right)}^{y}}^{*} \right)$, (3) draw $s_{\left( d \right)}^{v}$, (4) draw $s_{\left( d \right)}^{v}$ from a from a multinomial distribution, (5) draw $v_{\left( d \right)}$ from $f \left( v | z, \boldsymbol{c}_{\left( d \right)}; \omega_{l}^{v} \right)$; see \eqref{EDPM.eq.4.1.2}, (6) draw a sensitivity parameter $\rho_{\left( d \right)}$ from its prior, $f \left( \rho \right)$, (7) draw $v_{\left( d \right)}' $ from $\nP \left( V_{z'} = v' | V_{z} = v, Z = z, \boldsymbol{C} = \boldsymbol{c} \right)$; see \eqref{EDPM.eq.4.1.3}, (8) draw $s_{\left( d \right)}^{m'}$ from a multinomial distribution, and (9) draw $m_{\left( d \right)}'$ from $f \left( m' | v_{\left( d \right)}', z' , \boldsymbol{c}_{\left( d \right)}; \theta_{l}^{m*} \right)$. With $D$ values $\left( m_{\left( d \right)}', v_{\left( d \right)}', v_{\left( d \right)}, \boldsymbol{c}_{\left( d \right)}, s_{\left( d \right)}, \rho_{\left( d \right)} \right)$, the integral can be approximated as follows:
    \begin{align*}
        \mE{Y_{z, M_{z'}}} = \frac{1}{D} \sum_{d=1}^{D} \mE{Y | M = m_{\left( d \right)}', V = v_{\left( d \right)}, Z =z, \boldsymbol{C} = \boldsymbol{c}_{\left( d \right)}; \theta_{s_{\left( d \right)}^{y}}^{*} , \omega_{s_{\left( d \right)}^{\boldsymbol{x}}|s_{\left( d \right)}^{y}}^{*}, s_{\left( d \right)}, \rho_{\left( d \right)}}.
    \end{align*}
    where the conditional expectations in the sums are given in \eqref{EDPM.eq.4.1.1}. The steps above are performed after Gibbs sampling (details in \citet{roy2018bayesian}) since the steps above are all post MCMC. Missingness in $\left( Y, M, V, \boldsymbol{X} \right)$ where $\boldsymbol{X} = \left( Z, \boldsymbol{C}^{\top} \right)^{\top}$ is handled via data augmentation under an ignorable missingness assumption since the joint distribution of $\left( Y, M, V, \boldsymbol{X} \right)$ is specified; details in supplementary materials.
    
\section{Simulation} \label{EDPM.sec.5}
    To assess the performance of the proposed approach, we generated twelve data scenarios based on these introduced in \citet{kim2017framework, roy2018bayesian} and \citet{hong2022posttreatment}. The causal parameters of interest were the marginal natural direct effect, natural direct effect, and total effect. For the proposed Bayesian nonparametric (BNP) approach, we standardized the continuous covariates. We specified the following distributions in \eqref{EDPM.eq.5.1}: 
    \begin{align} \label{EDPM.eq.5.1}
        Y_{i} | M_{i}, V_{i}, \boldsymbol{X}_{i}; \theta_{i}^{y} & \sim \mnormals{ \mathbb{M}_{i} \beta_{i}^{y}, \sigma_{i}^{y,2} } \nonumber \\
        M_{i} | V_{i}, \boldsymbol{X}_{i}; \theta_{i}^{m} & \sim \mnormals{ \mathbb{V}_{i} \beta_{i}^{m}, \sigma_{i}^{m,2} } \nonumber \\
        V_{i} | \boldsymbol{X}_{i}; \omega_{i}^{v} & \sim \mnormals{ \mathbb{X}_{i} \beta_{i}^{v}, \sigma_{i}^{v,2} } \nonumber \\
        \boldsymbol{X}_{i,q}; \omega_{i,q}^{x} & \sim \mbernoulli{ \pi_{i,q}^{x} }, \; q=1, \cdots , 1 + p_{1} \\
        \boldsymbol{X}_{i,q}; \mu_{i,q}^{x}, \tau_{i,q}^{x,2} & \sim \mnormals{\mu_{i,q}^{x}, \tau_{i,q}^{x,2}}, \; q=1+p_{1}, \cdots ,1+p_{1}+p_{2} \nonumber \\
        \left( \theta_{i} , \omega_{i} \right) | \nG & \sim \nG \nonumber \\
        \nG & \sim \nedp \left( \alpha^{\theta}, \alpha^{\omega}, \nG_{0} \right) \nonumber 
    \end{align}
    where $\theta_{i}^{y} = \left( \beta_{i}^{y}, \sigma_{i}^{y,2} \right)$, $\theta_{i}^{m} = \left( \beta_{i}^{m}, \sigma_{i}^{m,2} \right)$, $\omega_{i}^{v} = \left( \beta_{i}^{v}, \sigma_{i}^{v,2} \right)$, $\omega_{i}^{x} = \left( \pi_{i}^{x}, \mu_{i}^{x}, \tau_{i}^{x,2} \right)$, $\theta_{i} = \left( \theta_{i}^{y}, \theta_{i}^{m} \right)$, and $\omega_{i} = \left( \omega_{i}^{v}, \omega_{i}^{x} \right)$. Here, $\mathbb{M}$, $\mathbb{V}$, and $\mathbb{X}$ are design matrices for each regression. For the prior specification, see the supplementary materials. We used a burn-in of $50,000$ and then used an additional $50,000$ samples for posterior inference for all twelve scenarios and sample sizes ($N=250, 500, 1000, 2500$). For the sensitivity parameter $\rho$, we specify a $\munif{0,1}$ prior. For each scenario, we generated 500 datasets and report the bias, mean squared error (MSE), the length of 95\% credible interval, and coverage probability. 

    We compared the EDPM model with the EDPM under the standard sequential ignorability assumption and a standard Bayesian parametric model under the extended version of the standard sequential ignorability assumption; for the former we dropped $V$ in \eqref{EDPM.eq.5.1} and for the latter we the standard Bayesian linear models were used for $Y$, $M$, and $V$, and a Bayesian bootstrap model was employed for the confounder distribution.
    \begin{align*}
        Y_{i} | M_{i}, V_{i}, \boldsymbol{X}_{i}; \theta_{i}^{y} & \sim \mnormals{ \mathbb{M}_{i} \beta_{i}^{y}, \sigma_{i}^{y,2} } \\
        M_{i} | V_{i}, \boldsymbol{X}_{i}; \theta_{i}^{m} & \sim \mnormals{ \mathbb{V}_{i} \beta_{i}^{m}, \sigma_{i}^{m,2} } \\
        V_{i} | \boldsymbol{X}_{i}; \theta_{i}^{v} & \sim \mnormals{ \mathbb{X}_{i} \beta_{i}^{v}, \sigma_{i}^{v,2} } \\
        \boldsymbol{X}_{i}; \omega_{i}^{x} & \sim \mbb{ \pi_{i}^{x} }
    \end{align*}
    where $\theta_{i}^{y} = \left( \beta_{i}^{y}, \sigma_{i}^{y,2} \right)$, $\theta_{i}^{m} = \left( \beta_{i}^{m}, \sigma_{i}^{m,2} \right)$, $\omega_{i}^{v} = \left( \beta_{i}^{v}, \sigma_{i}^{v,2} \right)$, $\omega_{i}^{x} = \left( \pi_{i}^{x}, \mu_{i}^{x}, \tau_{i}^{x,2} \right)$, $\theta_{i} = \left( \theta_{i}^{y}, \theta_{i}^{m} \right)$, and $\omega_{i} = \left( \omega_{i}^{v}, \omega_{i}^{x} \right)$. We used a similar priors with the EDPM specification. For the sensitivity parameter $\rho$, we specify a $\munif{0,1}$ prior.

    We generate a treatment variable from $Z \sim \mbernoulli{0.5}$ for all scenarios since our motivation example is a randomized trial. In scenarios 1-6, we generate continuous baseline confounders with simple functional forms; we generated two independent continuous baseline confounders $\left( C_{1}, C_{2} \right)$. We add more complex baseline confounders with complex functional forms in scenarios 7-12. In particular, we generated nine binary baseline confounders and six continuous baseline confounders, $\left( C_{1}, \cdots, C_{15} \right)$. We then sample a continuous post-treatment confounder, $\left( V_{z_{0}}, V_{z_{1}} \right)$, from a bivariate normal distribution in scenarios (1, 2, 3, 7, 8 and 9). In the rest of scenarios (4, 5, 6, 10, 11 and 12), a continuous post-treatment confounder is generated from a bivariate gamma distribution. For each scenario, we generate a mediator, $M$, from a skewed normal distribution. An outcome, $Y$, is generated from a mixture of normals except for scenarios 3, 6, 9 and 12. For an outcome in scenarios 3, 6, 9 and 12, we add more complex terms to the mean of the data generation mechanism. See the Table \ref{EDPM:tab:sim:scn} for detailed specifications.
    
\subsection{Results} \label{EDPM.subsec.5.1}
    Table \ref{EDPM:tab:sim:wV} presents the results of the simulation study. Overall, the results show that all of the estimates are approximately unbiased even when the sample size is low ($N=250$). As the sample size increases, the absolute value of bias, MSE, and the length of the 95\% credible interval decrease, while the empirical coverage probability remains close to 95\%.

    The proposed model is robust to violations of the bivariate normality assumptions for post-treatment confounders. This can be seen by comparing the results of scenarios 1 and 4, scenarios 2 and 5, scenarios 3 and 6, scenarios 7 and 10, and scenarios 8 and 11, and scenarios 9 and 12. However, the MSE and the length of the 95\% credible interval for scenarios 4, 5, and 6 tend to be slightly smaller compared to scenarios 1, 2, and 3, respectively. Furthermore, the MSE and the length of the 95\% credible interval are bigger when there is an interaction term in the mean of normals (scenarios 1 vs 2, scenarios 4 vs 5, scenarios 7 vs 8, and scenarios 10 vs 11). In the setting of nonlinear terms (scenarios 3, 6, 9, 12), the absolute values of bias for NIE and NDE are slightly larger, but the bias vanishes as the sample size increases, as expected.

    The simulation results are presented in Table \ref{EDPM:tab:sim:woV} with the EDPM under the standard sequential ignorability assumption where we dropped $V$ in \eqref{EDPM.eq.5.1}. Overall, the results suggest that most of the estimates are biased and that MSE and the length of the 95\% credible interval increase even when the sample size increases. Additionally, we observed that the empirical coverage probability deviates significantly from the target value of 95\% for many scenarios.

    Table \ref{EDPM:tab:sim:para} presents the results of the simulation study with a standard Bayesian parametric model under the extended sequential ignorability assumption. The results suggest that the estimates are mostly unbiased for all cases except for case 3, and that MSE and the length of the 95\% credible interval decrease as sample size increases. However, for at least one of the causal effects (NIE, NDE, or ATE) for each scenario, the empirical coverage probability is less than 95\% (sometimes by a very large amount, see e.g., S3). This suggests that the Bayesian parametric model tends to underestimate the length of the credible interval. 
    
\section{Rural LITE Trial} \label{EDPM.sec.6}
    We used the proposed method to assess mediation in the Rural LITE trial \citep{perri2014comparative}. The Rural LITE trial was designed to examine the effects and costs of behavioral weight-loss treatment delivered via Cooperative Extension Offices in rural communities. subjects ($N = 612$) were randomized to one of 4 treatment arms: (1) control ($N = 169$); (2) low dose ($N = 148$); (3) moderate dose ($N = 134$); (4) high dose ($N = 161$). There was interest in exploring the attendance rate as a potential mediator of weight change during the study. Weight was measured at months 0 (baseline), 6, and 24. The primary outcome is change in body weight from the baseline to the final follow-up (month 24). The weight change from the baseline to month 6 is regarded as a post-treatment confounder. We included baseline confounders sex, age, age, race, and BMI.
    
    In the analysis of the original trial, the control and low dose face treatment arms resulted in similar weight maintenance. Thus, we combine these interventions into a single intervention group ($z=0$). For the same reason, we combine the moderate dose and high dose ($z=1$). Here, we assess the NIE and NDE of the mediator for the lower dose (LD) versus higher dose (HD) arms. The sample sizes for the two treatment arms were $N = 317$ and $N = 295$, respectively. There were 121 subjects (weight at month 24) and 56 (weight at month 6) subjects who had incomplete information on the primary outcome and/or a post-treatment confounder, respectively. We handled these under ignorable missingness via a Monte Carlo Markov chain (MCMC) data augmentation algorithm conditional on observed data and parameters in the EDPM; see the supplementary materials for details. For the sensitivity analysis for a post-treatment confounder assumption \eqref{EDPM.eq.3.4}, we examined multiple values for the sensitivity parameter $\rho$ as well as several priors.
    
\subsection{Results} \label{EDPM.subsec.6.1}
    For the prior distributions for the observed data model parameters, we use the same specification as in the simulation study. For sampling from the posterior distribution for the observed data, we ran 100,000 iterations and discarded the first 50,000 as burn-in with thinning of 100 to obtain 500 posterior samples.

   Table \ref{EDPM:tab:avgmain} presents the estimates of NIE, NDE, and ATE under various choices of sensitivity parameter $\rho$ based on the assumptions described in Section \ref{EDPM.sec.3}. The results suggest that the proposed model is robust to the choice of $\rho$ for ATE and NDE estimates, which have a significant effect, indicated by the 95\% credible intervals excluding zero. On the other hand, the NIE estimate yield results that mostly vary depending on the sign of $\rho$. In particular, NIE is statistically significant ($i.e.$, 95\% credible interval excluding 0) for $\rho < 0$ while NIE is not statistically significant for $\rho > 0$. The ATE estimates for different values of $\rho$ are similar as expected, indicating that ATE is insensitive to the choice of $\rho$ (as it should be). 
    
    The NDE represents the average difference in weight change between the HD ($z=1$) and LD ($z=0$) groups at the end of the study, if the attendance of the HD group was set to what it would have been without the intervention. A negative and statistically significant NDE suggests that the HD has a significant effect on weight change that is not due to attendance. On the other hand, the NIE represents the average weight change if all subjects were in the HD arm, compared to the average weight change if they were also in the HD arm but their attendance was set to what it would have been without the intervention. The fact that the 95\% CIs of the NIEs in Table \ref{EDPM:tab:avgmain} include zero when $\rho > 0$ and exclude $\rho < 0$ suggests that the effects of the HD intervention that are due to changes in attendance are not significant when $\rho > 0$ and are significant when $\rho < 0$, respectively. Our analysis shows that the HD intervention is more effective in terms of average weight loss. There is conflicting evidence regarding whether the effect of the HD intervention is mediated by attendance rate or not during the study.

    We also considered causal effects conditional on race (Black or White); computations for this are in section {{3}} of the supplementary materials. Table \ref{EDPM:tab:condmain} contains the estimates of NIE, NDE, and TE conditional on race for different values of the sensitivity parameter $\rho$. The results are consistent across values of $\rho$ for race. The estimates of ATE for blacks are lower than those for whites. The ATE and NDE are significant ($i.e.$, 95\% credible intervals exclude 0) for both races. The NIEs are not significant ($i.e.$, 95\% credible intervals include 0). However, the NIE estimates for whites are almost significant, with the upper bounds of the 95\% credible intervals being very close to 0.
    
    We also performed additional data analysis to examine overall mediation effects with different $\rho$ values (Table {{1}} in the supplementary materials) and under the standard sequential ignorability assumption, where the post-treatment confounder was dropped (Table {{2}} in the supplementary materials). Notably, we found little evidence of NIE (centered at 0) in the absence of a post-treatment confounder. Furthermore, we conducted additional data analysis to investigate conditional mediation effects with different $\rho$ values, and the results are presented in Table {{3}} in the supplementary materials.
    
\section{Discussion} \label{EDPM.sec.7}
    This article proposed a flexible BNP approach for the causal effects of mediation in the presence of a post-treatment confounder. In particular, we use EDPM models for the observed data combined with a recently proposed set of assumptions for identification in the presence of a post-treatment confounder. Our BNP method for estimating causal effects provides greater flexibility than parametric methods, while still maintaining computational ease, making it a valuable tool for causal inference research.
    
    For the Rural LITE data, we found that the effect of the HD intervention and NDE were statistically significant, while there was conflicting evidence regarding whether the effect of the HD intervention is mediated by attendance rate or not. Additionally, we observed that the posteriors of the indirect and direct effects were relatively insensitive to non-negative $\rho$ values. Moreover, we found that the results obtained by assuming standard sequential ignorability without including a post-treatment confounder estimated larger NDE and ATE and no evidence of NIE. 
    
    Future work will explore extensions of the proposed model to allow for the inclusion of multiple post-treatment confounders and/or mediators. Additionally, sensitivity analysis will be important to assess potential violations of assumptions \eqref{EDPM.eq.3.2} and/or \eqref{EDPM.eq.3.3}. One potential direction for identification in the presence of a post-treatment confounder is to explore weaker assumptions such as a mixture of mediator induction equivalence assumptions as proposed in \citep{daniels2012bayesian, kim2017framework}, and the assumption proposed in \citet{hong2022posttreatment}.
    
\section*{Supporting Information}
    Web appendices, Tables, and Figures (which further describe model specification, posterior computation, simulation set-up, missing handling) referenced in Sections {{1}}, {{2}}, {{3}}, {{4}}, {{5}}, and {{6}} along with the code are available with this article at \url{https://github.com/WoojungBae/EDPpostMediation}.
    
\section*{Acknowledgements}
    This work was supported by the following NIH grants: R01CA183854, R01HL158963, R01HL166324.

    {
    \linespread{1}\selectfont
    \bibliographystyle{agu}
    \bibliography{fileReference}      
    }
    
\newpage
    \begin{figure}[htbp]
    \centering
    \begin{tikzpicture} 
        [
            rect/.style={minimum size=30pt,rectangle,draw},
            circ/.style={minimum size=30pt,circle,draw},
            thickarro/.style={->,>=stealth',semithick},
            arro/.style={->,>=stealth',},
            dottarro/.style={->,>=stealth',dotted},
            dasharro/.style={->,>=stealth',dashed},
        ]
        
        \node (Z) [circ] at (-6,0) {$Z$};
        \node (V) [circ] at (-2,0) {$V$};
        \node (M) [circ] at (2,0) {$M$};
        \node (Y) [circ] at (6,0) {$Y$};
        
        \node (C) [circ] at (0,-2) {$\boldsymbol{C}$};
        
        \draw[dasharro] (Z) -- (V) node at (-4,-0.3) {};
        \draw[dasharro] (V) -- (M) node at (0,-0.3) {};
        \draw[thickarro] (M) -- (Y) node at (4,-0.3) {};
        
        \draw[dottarro] (C) -- (V) node at (-1.4,-1) {};
        \draw[dottarro] (C) -- (M) node at (1.4,-1) {};
        \draw[dottarro] (C) -- (Y) node at (3,-1.3) {};
        
        
        
        
        \path [dasharro,bend left] (V) edge (Y);
        \path [arro,bend right] (Z) edge (M);
        \path [arro,bend left] (Z) edge (Y);
    \end{tikzpicture}
    \caption{\label{EDPM.ce.post} The causal structure with a post-treatment confounder, $V$ and a randomized treatment $Z$ where $Y$ is an outcome, $M$ is a mediator and $C$ is a set of baseline confounders.}
\end{figure}
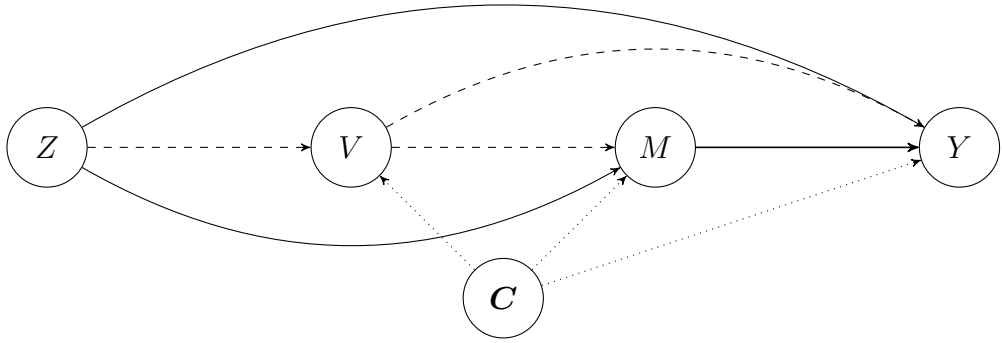 
    \begin{table}[htbp] 
    \fontsize{6}{14}\selectfont
    \caption{\label{EDPM:tab:sim:scn} Data generating mechanism for each simulation scenario}
    \centering
        \begin{tabular}[htbp]{c|c|l}
        \toprule
         & Scn & \multicolumn{1}{c}{Data Generation} \\
        
        \midrule
        \midrule
        $C_{1}$ & \multirow{2}{*}{1--6} & $ \sim \mnormals{\mu_{1}, \sigma_{1}^{2}} $; $\mu_{1} = 0$, $\sigma_{1}^{2} = 9$\\
        $C_{2}$ &  & $ \sim \mnormals{\mu_{2}, \sigma_{2}^{2}} $; $\mu_{2} = 0$, $\sigma_{2}^{2} = 16$ \\
        
        \cmidrule{1-3}
        \multirow{4}{*}{ $\begin{pmatrix} V_{z_{0}} \\ V_{z_{1}} \end{pmatrix} $ } & 1--3 & $ \sim \mnormals{ \begin{pmatrix} \mu_{0} \\ \mu_{1} \end{pmatrix} , \begin{pmatrix} \sigma_{0}^{2} & \rho_{01} \sigma_{0} \sigma_{1} \\ \rho_{01} \sigma_{0} \sigma_{1} & \sigma_{1}^{2} \end{pmatrix} ; \rho_{01} } $ \\
        & 4--6 & $ \sim \text{BVG} \left( \begin{pmatrix} \mgamma{\text{shape} = \log \left( 1 + \exp \left( \mu_{0} \right) \right) , \text{rate} = 1} \\ \mgamma{\text{shape} = \log \left( 1 + \exp \left( \mu_{1} \right) \right) , \text{rate} = 1}\end{pmatrix} ; \rho_{01} \right) $ \\
        & & \hspace{1em} where $\mu_{0} = 1.3 + 0.6 C_{1} - 0.7 C_{2}$, $\mu_{1} = 1.4 - 0.5 C_{1} + 0.3 C_{2}$, $\sigma_{0}^{2} = 3$, $\sigma_{1}^{2} = 10$, $\rho_{01} = 0.3$ \\
         
        \cmidrule{1-3}
        \multirow{3}{*}{M} & 1 \& 4 & $ \sim \mskewednormals{\xi, \omega, \alpha} $, $\xi = 1.0 + 1.7 Z + 0.5 V + 0.4 C_{1} + 0.9 C_{2}$, $\omega = 3$, $\alpha = 10$ \\
        & 2 \& 5 & $ \sim \mskewednormals{\xi, \omega, \alpha} $, $\xi = 1.0 + 1.7 Z + 0.5 V + 0.4 C_{1} + 0.3 C_{2}$, $\omega = 3$, $\alpha = 10$ \\
        & 3 \& 6 & $ \sim \mskewednormals{\xi, \omega, \alpha} $, $\xi = -1.5 + 0.5 Z + 0.1 V + 0.1 C_{1} + 0.3 C_{2}$, $\omega = 1$, $\alpha = 7$ \\
        
        \cmidrule{1-3}
        \multirow{8}{*}{Y} & \multirow{3}{*}{1 \& 4} & $ \sim 0.6 \mnormals{\mu_{1}, \sigma_{1}^{2}} +  0.4 \mnormals{\mu_{2}, \sigma_{2}^{2}} $ \\
         & & \hspace{1em} where $\mu_{1} = 5 + 2.5 Z + 1.8 M + 1.3 V - 1.2 C_{1} + 0.3 C_{2}$, $\sigma_{1}^{2} = 1.5$ \\
         & & \hspace{1em} where $\mu_{2} = - 5 - 1.5 Z - 1.0 M - 0.7 V + 0.4 C_{1} + 0.3 C_{2}$, $\sigma_{2}^{2} = 0.5$ \\
         & \multirow{3}{*}{2 \& 5} & $ \sim 0.6 \mnormals{\mu_{1}, \sigma_{1}^{2}} +  0.4 \mnormals{\mu_{2}, \sigma_{2}^{2}} $  \\
         & & \hspace{1em} where $\mu_{1} = 5 + 2.5 Z + 1.8 M + 1.0 Z \times M + 1.3 V - 0.4 C_{1} + 0.3 C_{2}$, $\sigma_{1}^{2} = 1.5$ \\
         & & \hspace{1em} where $\mu_{2} = - 5 - 1.5 Z - 1.0 M - 0.5 Z \times M - 0.7 V + 0.4 C_{1} + 0.3 C_{2}$, $\sigma_{2}^{2} = 0.5$ \\
         & \multirow{2}{*}{3 \& 6} & $ \mnormals{ \mu , \sigma^{2}} $ \\
         & & \hspace{1em} where $\mu = 5 + 2.5 Z + 0.2 \left( M - 0.4 \right)_{+} + 0.6 M^{2} + 0.3 V + 0.4 C_{1} + 0.3 C_{2}$, $\sigma^{2} = 0.2$ \\
         
        \midrule
        \midrule
        $C_{1} \cdots C_{3}$ & \multirow{5}{*}{7--12} & $ \sim \mbernoulli{\pi} $; $\pi = 0.05$ \\
        $C_{4} \cdots C_{6}$ &  & $ \sim \mbernoulli{\pi} $; $\pi = 0.5$ \\
        $C_{7} \cdots C_{9}$ &  & $ \sim \mbernoulli{\pi_{C}} $; $\pi_{C} = \minvlogit{a_{C} \minvlogit{b_{C}} + \left( 1 - a_{C} \right) \minvlogit{c_{C}}}$ \\
         & & \hspace{1em} where $a_{C} = \minvlogit{2 \left( C_{10} - 2 \right)^{2} - 2 \left( C_{11} + 1 \right)^{2}}$, $b_{C} = 0.6 C_{12} C_{13} - 0.2 C_{14}^{2}$, $c_{C} = 0.7 C_{12} - 0.4 C_{14} C_{15}$ \\
        $C_{10} \cdots C_{15}$ &  & $ \sim \mmvns{\pmb{0} , 0.7 \pmb{I} + 0.3 \pmb{1} \pmb{1}^{\top}}$, $\mu = \pmb{0}$, $\Sigma = 0.7 \pmb{I} + 0.3 \pmb{1} \pmb{1}^{\top}$ \\
        
        \cmidrule{1-3}
        \multirow{5}{*}{ $\begin{pmatrix} V_{z_{0}} \\ V_{z_{1}} \end{pmatrix} $ } & 1--3 & $ \sim \mnormals{ \begin{pmatrix} \mu_{0} \\ \mu_{1} \end{pmatrix} , \begin{pmatrix} \sigma_{0}^{2} & \rho_{01} \sigma_{0} \sigma_{1} \\ \rho_{01} \sigma_{0} \sigma_{1} & \sigma_{1}^{2} \end{pmatrix} ; \rho_{01} } $ \\
        & 4--6 & $ \sim \text{BVG} \left( \begin{pmatrix} \mgamma{\text{shape} = \log \left( 1 + \exp \left( \mu_{0} \right) \right) , \text{rate} = 1} \\ \mgamma{\text{shape} = \log \left( 1 + \exp \left( \mu_{1} \right) \right) , \text{rate} = 1}\end{pmatrix} ; \rho_{01} \right) $ \\
         & & \hspace{1em} where $\mu_{0} = 1.3 + 0.5 C_{10} - 0.7 C_{11} + 0.3 C_{12}$, $\mu_{1} = 1.4 + 0.3 C_{13} - 0.2 C_{14} - 0.4 C_{15}$, $\sigma_{0}^{2} = 3$, $\sigma_{1}^{2} = 10$, $\rho_{01} = 0.3$ \\
         
        \cmidrule{1-3}
        \multirow{3}{*}{M} & 7 \& 10 & $ \sim \mskewednormals{\xi, \omega, \alpha} $, $\xi = -1.0 + 1.7 Z + 0.5 V + 0.3 C_{1} + 0.1 C_{4} - 0.2 C_{7} - 0.4 C_{10} + 0.6 C_{13}$, $\omega = 3$, $\alpha = 10$ \\
        & 8 \& 11 & $ \sim \mskewednormals{\xi, \omega, \alpha} $, $\xi = -1.0 + 1.7 Z + 0.5 V + 0.3 C_{1} + 0.1 C_{4} - 0.2 C_{7} - 0.4 C_{10} + 0.6 C_{13}$, $\omega = 3$, $\alpha = 10$ \\
        & 9 \& 12 & $ \sim \mskewednormals{\xi, \omega, \alpha} $, $\xi = -0.7 + 0.2 Z + 0.1 V + 0.5 C_{1} + 0.6 C_{4} - 0.2 C_{7} - 0.4 C_{10} + 0.6 C_{13}$, $\omega = 1$, $\alpha = 7$ \\
        
        \cmidrule{1-3}
        \multirow{8}{*}{Y} & \multirow{3}{*}{7 \& 10} & $ \sim 0.6 \mnormals{\mu_{1}, \sigma_{1}^{2}} +  0.4 \mnormals{\mu_{2}, \sigma_{2}^{2}} $ \\
         & & \hspace{1em} where $\mu_{1} = 5 + 2.5 Z + 1.8 M + 1.3 V + 0.1 C_{2} + 0.3 C_{5} - 0.4 C_{8} - 0.2 C_{11} + 0.6 C_{14}$, $\sigma_{1}^{2} = 1.5$ \\
         & & \hspace{1em} where $\mu_{2} = -5 - 0.5 Z - 1.0 M - 0.7 V + 0.3 C_{3} + 0.1 C_{6} - 0.2 C_{9} + 0.6 C_{12} - 0.4 C_{15}$, $\sigma_{2}^{2} = 0.5$ \\
         & \multirow{3}{*}{8 \& 11} & $ \sim 0.6 \mnormals{\mu_{1}, \sigma_{1}^{2}} +  0.4 \mnormals{\mu_{2}, \sigma_{2}^{2}} $  \\
         & & \hspace{1em} where $\mu_{1} = 5 + 2.5 Z + 1.8 M + 1.0 Z * M + 1.3 V + 0.1 C_{2} + 0.3 C_{5} - 0.4 C_{8} - 0.2 C_{11} + 0.6 C_{14}$, $\sigma_{1}^{2} = 1.5$ \\
         & & \hspace{1em} where $\mu_{2} = -5 - 1.5 Z - 1.0 M - 1.5 Z * M + 0.7 V + 0.3 C_{3} + 0.1 C_{6} - 0.2 C_{9} + 0.6 C_{12} - 0.4 C_{15}$, $\sigma_{2}^{2} = 0.5$ \\
         & \multirow{2}{*}{9 \& 12} & $ \sim \mnormals{ \mu , \sigma^{2}} $ \\
         & & \hspace{1em} where $\mu = 5 + 2.5 Z + 0.2 \left( M - 0.4 \right)_{+} + 0.6 M^{2} + 0.3 V + 0.2 C_{2} + 0.3 C_{5} - 0.4 C_{8} - 0.2 C_{11} + 0.6 C_{14}$, $\sigma^{2} = 0.2$ \\
        \bottomrule
    \end{tabular}
\end{table} 
    \begin{table}[htbp]
    \fontsize{6}{11}\selectfont
    \centering
    \caption{\label{EDPM:tab:sim:wV} Simulation results for NIE, NDE, and ATE over 500 replications. The columns correspond to bias, MSE, CIl (length of 95\% credible interval), and CP (the empirical coverage probability).}
    \begin{tabular}[htbp]{llr|rrrr|rrrr|rrrr|rrrr} 
        \toprule
        \multicolumn{3}{c}{ } & \multicolumn{4}{c}{250} & \multicolumn{4}{c}{500} & \multicolumn{4}{c}{1000} & \multicolumn{4}{c}{2500} \\
        \cmidrule(l{3pt}r{3pt}){4-7} \cmidrule(l{3pt}r{3pt}){8-11} \cmidrule(l{3pt}r{3pt}){12-15} \cmidrule(l{3pt}r{3pt}){16-19}
        Scn & CE & Truth & Bias & MSE & CIl & CP & Bias & MSE & CIl & CP & Bias & MSE & CIl & CP & Bias & MSE & CIl & CP \\
        \midrule
        & NIE & 1.19 & 0.01 & 4.54 & 8.99 & 0.97 & 0.07 & 2.52 & 7.31 & 0.98 & -0.10 & 1.73 & 5.83 & 0.97 & 0.02 & 0.91 & 4.15 & 0.98\\
        
        & NDE & 0.95 & -0.13 & 6.34 & 9.62 & 0.95 & 0.03 & 3.27 & 7.81 & 0.98 & 0.04 & 2.30 & 6.37 & 0.96 & -0.03 & 1.04 & 5.02 & 0.98\\
        
        \multirow{-3}{*}{S1} & ATE & 2.14 & 0.12 & 4.33 & 6.71 & 0.90 & 0.06 & 2.18 & 4.98 & 0.91 & -0.08 & 1.13 & 4.20 & 0.95 & 0.00 & 0.41 & 3.76 & 1.00\\
        \cmidrule{1-19}
        & NIE & 1.89 & 0.08 & 5.03 & 9.72 & 0.97 & 0.05 & 2.71 & 7.82 & 0.98 & -0.08 & 1.89 & 6.19 & 0.98 & 0.00 & 1.05 & 4.41 & 0.97\\
        
        & NDE & 2.56 & -0.11 & 7.41 & 10.28 & 0.95 & 0.08 & 3.71 & 8.30 & 0.96 & 0.05 & 2.52 & 6.74 & 0.96 & 0.00 & 1.13 & 5.42 & 0.98\\
        
        \multirow{-3}{*}{S2} & ATE & 4.45 & 0.10 & 5.89 & 7.61 & 0.90 & 0.13 & 2.96 & 5.64 & 0.91 & -0.06 & 1.47 & 4.80 & 0.95 & 0.04 & 0.53 & 4.40 & 1.00\\
        \cmidrule{1-19}
        & NIE & 0.30 & -0.05 & 0.03 & 0.86 & 0.98 & -0.02 & 0.01 & 0.63 & 0.98 & 0.02 & 0.01 & 0.51 & 0.98 & 0.04 & 0.01 & 0.42 & 1.00\\
        
        & NDE & 2.53 & -0.05 & 0.04 & 1.31 & 1.00 & -0.05 & 0.02 & 0.87 & 0.99 & -0.05 & 0.01 & 0.64 & 0.99 & -0.05 & 0.01 & 0.48 & 0.99\\
        
        \multirow{-3}{*}{S3} & ATE & 2.83 & -0.07 & 0.05 & 1.13 & 0.99 & -0.05 & 0.03 & 0.78 & 0.97 & -0.02 & 0.01 & 0.59 & 0.99 & -0.01 & 0.00 & 0.49 & 1.00\\
        \cmidrule{1-19}
        & NIE & 1.04 & 0.06 & 2.03 & 6.23 & 0.97 & 0.00 & 1.12 & 5.04 & 0.98 & -0.03 & 0.86 & 4.02 & 0.98 & -0.06 & 0.42 & 2.85 & 0.97\\
        
        & NDE & 0.72 & -0.02 & 4.80 & 8.14 & 0.95 & 0.16 & 2.49 & 6.38 & 0.97 & 0.05 & 1.59 & 5.24 & 0.96 & -0.01 & 0.71 & 4.37 & 0.99\\
        
        \multirow{-3}{*}{S4} & ATE & 1.76 & 0.08 & 4.42 & 7.05 & 0.91 & 0.16 & 2.27 & 5.25 & 0.92 & 0.03 & 1.18 & 4.32 & 0.94 & 0.02 & 0.44 & 3.99 & 1.00\\
        \cmidrule{1-19}
        & NIE & 1.64 & 0.04 & 2.32 & 6.69 & 0.97 & 0.04 & 1.15 & 5.29 & 0.99 & -0.04 & 0.89 & 4.17 & 0.97 & -0.05 & 0.42 & 2.96 & 0.98\\
        
        & NDE & 2.52 & 0.00 & 5.35 & 8.45 & 0.95 & 0.10 & 2.73 & 6.61 & 0.95 & 0.01 & 1.73 & 5.49 & 0.96 & 0.03 & 0.73 & 4.71 & 1.00\\
        
        \multirow{-3}{*}{S5} & ATE & 4.17 & 0.12 & 5.57 & 7.89 & 0.91 & 0.21 & 2.85 & 5.86 & 0.92 & -0.01 & 1.51 & 4.93 & 0.96 & 0.01 & 0.54 & 4.57 & 1.00\\
        \cmidrule{1-19}
        & NIE & 0.18 & 0.03 & 0.02 & 0.69 & 0.99 & 0.01 & 0.01 & 0.51 & 0.99 & 0.02 & 0.01 & 0.42 & 0.99 & 0.03 & 0.00 & 0.34 & 1.00\\
        
        & NDE & 2.39 & -0.09 & 0.03 & 1.17 & 1.00 & -0.07 & 0.02 & 0.76 & 0.99 & -0.05 & 0.01 & 0.54 & 1.00 & -0.05 & 0.00 & 0.39 & 1.00\\
        
        \multirow{-3}{*}{S6} & ATE & 2.57 & -0.04 & 0.03 & 1.07 & 0.99 & -0.05 & 0.02 & 0.73 & 0.98 & -0.03 & 0.01 & 0.55 & 0.99 & -0.02 & 0.00 & 0.44 & 1.00\\
        \cmidrule{1-19}
        & NIE & 1.19 & -0.01 & 1.37 & 4.18 & 0.92 & -0.05 & 0.51 & 3.00 & 0.94 & 0.02 & 0.29 & 2.30 & 0.95 & -0.03 & 0.14 & 1.67 & 0.96\\
        
        & NDE & 0.95 & 0.06 & 2.74 & 7.59 & 0.96 & 0.00 & 0.94 & 4.19 & 0.97 & -0.02 & 0.63 & 3.20 & 0.97 & 0.01 & 0.26 & 2.30 & 0.97\\
        
        \multirow{-3}{*}{S7} & ATE & 2.14 & 0.15 & 2.44 & 6.70 & 0.96 & 0.03 & 1.15 & 4.28 & 0.96 & -0.01 & 0.59 & 3.15 & 0.95 & -0.02 & 0.24 & 2.25 & 0.98\\
        \cmidrule{1-19}
        & NIE & 1.89 & -0.10 & 2.13 & 4.93 & 0.89 & -0.01 & 0.77 & 3.52 & 0.95 & -0.01 & 0.42 & 2.74 & 0.96 & -0.03 & 0.20 & 1.97 & 0.96\\
        
        & NDE & 1.74 & 0.08 & 3.59 & 8.61 & 0.96 & -0.01 & 1.14 & 4.46 & 0.97 & 0.02 & 0.72 & 3.44 & 0.97 & -0.02 & 0.30 & 2.48 & 0.97\\
        
        \multirow{-3}{*}{S8} & ATE & 3.63 & 0.06 & 3.22 & 7.64 & 0.96 & 0.06 & 1.57 & 4.97 & 0.94 & 0.02 & 0.81 & 3.64 & 0.96 & -0.03 & 0.33 & 2.64 & 0.98\\
        \cmidrule{1-19}
        & NIE & 0.30 & -0.04 & 0.02 & 0.48 & 0.91 & -0.02 & 0.01 & 0.36 & 0.93 & -0.02 & 0.00 & 0.28 & 0.95 & -0.02 & 0.00 & 0.22 & 0.97\\
        
        & NDE & 2.51 & -0.03 & 0.01 & 0.59 & 0.99 & -0.04 & 0.01 & 0.43 & 0.96 & -0.05 & 0.01 & 0.33 & 0.98 & -0.06 & 0.00 & 0.27 & 0.98\\
        
        \multirow{-3}{*}{S9} & ATE & 2.81 & -0.06 & 0.04 & 0.81 & 0.97 & -0.07 & 0.02 & 0.57 & 0.94 & -0.07 & 0.01 & 0.44 & 0.93 & -0.07 & 0.01 & 0.34 & 0.94\\
        \cmidrule{1-19}
        & NIE & 1.17 & -0.02 & 1.46 & 4.10 & 0.89 & -0.03 & 0.49 & 2.58 & 0.93 & 0.01 & 0.21 & 1.85 & 0.95 & 0.00 & 0.10 & 1.28 & 0.96\\
        
        & NDE & 0.92 & 0.09 & 3.10 & 7.67 & 0.96 & 0.07 & 1.08 & 3.97 & 0.96 & 0.00 & 0.53 & 2.89 & 0.95 & 0.00 & 0.22 & 2.02 & 0.97\\
        
        \multirow{-3}{*}{S10} & ATE & 2.10 & 0.12 & 2.47 & 6.62 & 0.96 & 0.04 & 1.05 & 4.12 & 0.94 & -0.01 & 0.57 & 3.00 & 0.95 & 0.03 & 0.21 & 2.06 & 0.98\\
        \cmidrule{1-19}
        & NIE & 1.86 & -0.04 & 2.15 & 4.62 & 0.89 & -0.05 & 0.61 & 3.01 & 0.94 & 0.02 & 0.31 & 2.21 & 0.95 & 0.00 & 0.15 & 1.55 & 0.95\\
        
        & NDE & 1.77 & 0.03 & 3.69 & 8.57 & 0.96 & 0.06 & 1.19 & 4.13 & 0.96 & 0.00 & 0.66 & 3.11 & 0.96 & 0.01 & 0.27 & 2.17 & 0.96\\
        
        \multirow{-3}{*}{S11} & ATE & 3.63 & 0.08 & 3.07 & 7.46 & 0.95 & 0.00 & 1.43 & 4.69 & 0.94 & 0.01 & 0.77 & 3.49 & 0.95 & 0.03 & 0.28 & 2.42 & 0.97\\
        \cmidrule{1-19}
        & NIE & 0.26 & 0.01 & 0.01 & 0.46 & 0.95 & 0.02 & 0.01 & 0.34 & 0.92 & 0.01 & 0.00 & 0.27 & 0.95 & 0.00 & 0.00 & 0.22 & 0.99\\
        
        & NDE & 2.50 & -0.06 & 0.01 & 0.42 & 0.98 & -0.08 & 0.01 & 0.30 & 0.91 & -0.07 & 0.01 & 0.23 & 0.92 & -0.05 & 0.00 & 0.20 & 0.96\\
        
        \multirow{-3}{*}{S12} & ATE & 2.76 & -0.06 & 0.02 & 0.61 & 0.95 & -0.06 & 0.01 & 0.43 & 0.94 & -0.06 & 0.01 & 0.32 & 0.93 & -0.05 & 0.00 & 0.25 & 0.95\\
        \bottomrule
    \end{tabular}
\end{table} 
    \begin{table}[htbp]
    \fontsize{6}{11}\selectfont
    \centering
    \caption{\label{EDPM:tab:sim:woV} Simulation results for NIE, NDE, and ATE without a post-treatment variable over 500 replications. The columns correspond to bias, MSE, CIl (length of 95\% credible interval), and CP (the empirical coverage probability).}
    \begin{tabular}[htbp]{llr|rrrr|rrrr|rrrr|rrrr} 
        \toprule
        \multicolumn{3}{c}{ } & \multicolumn{4}{c}{250} & \multicolumn{4}{c}{500} & \multicolumn{4}{c}{1000} & \multicolumn{4}{c}{2500} \\
        \cmidrule(l{3pt}r{3pt}){4-7} \cmidrule(l{3pt}r{3pt}){8-11} \cmidrule(l{3pt}r{3pt}){12-15} \cmidrule(l{3pt}r{3pt}){16-19}
        Scn & CE & Truth & Bias & MSE & CIl & CP & Bias & MSE & CIl & CP & Bias & MSE & CIl & CP & Bias & MSE & CIl & CP \\
        \midrule
        & NIE & 1.19 & 0.90 & 1.93 & 4.06 & 0.86 & 0.88 & 1.25 & 3.17 & 0.82 & 0.86 & 0.99 & 2.90 & 0.81 & 1.05 & 1.28 & 3.51 & 0.85\\
        
        & NDE & 0.95 & -0.89 & 4.42 & 8.30 & 0.96 & -0.79 & 2.21 & 5.74 & 0.92 & -0.92 & 1.67 & 4.28 & 0.89 & -1.10 & 1.54 & 4.04 & 0.91\\
        
        \multirow{-3}{*}{S1} & ATE & 2.14 & 0.12 & 4.62 & 8.05 & 0.93 & 0.14 & 2.24 & 5.56 & 0.94 & -0.02 & 1.11 & 4.07 & 0.95 & 0.00 & 0.39 & 3.11 & 0.98\\
        \cmidrule{1-19}
        & NIE & 1.89 & 0.80 & 3.03 & 5.36 & 0.88 & 0.87 & 1.72 & 3.79 & 0.84 & 0.87 & 1.17 & 2.97 & 0.79 & 0.99 & 1.21 & 2.99 & 0.75\\
        
        & NDE & 2.56 & -0.71 & 5.16 & 9.75 & 0.96 & -0.84 & 2.72 & 6.61 & 0.95 & -0.97 & 1.96 & 4.59 & 0.89 & -1.02 & 1.44 & 3.76 & 0.90\\
        
        \multirow{-3}{*}{S2} & ATE & 4.45 & 0.09 & 5.98 & 9.27 & 0.93 & 0.10 & 2.93 & 6.43 & 0.95 & -0.03 & 1.46 & 4.66 & 0.94 & 0.02 & 0.51 & 3.68 & 0.99\\
        \cmidrule{1-19}
        & NIE & 0.30 & 0.34 & 0.14 & 0.74 & 0.53 & 0.36 & 0.15 & 0.60 & 0.26 & 0.40 & 0.17 & 0.49 & 0.03 & 0.42 & 0.18 & 0.39 & 0.00\\
        
        & NDE & 2.53 & -0.42 & 0.20 & 0.95 & 0.64 & -0.42 & 0.19 & 0.64 & 0.24 & -0.43 & 0.19 & 0.45 & 0.05 & -0.43 & 0.19 & 0.31 & 0.00\\
        
        \multirow{-3}{*}{S3} & ATE & 2.83 & -0.08 & 0.05 & 0.94 & 0.97 & -0.05 & 0.03 & 0.67 & 0.96 & -0.02 & 0.01 & 0.52 & 0.98 & -0.01 & 0.00 & 0.43 & 1.00\\
        \cmidrule{1-19}
        & NIE & 1.04 & 0.47 & 1.32 & 4.09 & 0.92 & 0.48 & 0.72 & 3.33 & 0.93 & 0.79 & 1.04 & 3.74 & 0.92 & 1.01 & 1.55 & 3.57 & 0.90\\
        
        & NDE & 0.72 & -0.40 & 4.31 & 8.54 & 0.96 & -0.35 & 1.96 & 5.86 & 0.95 & -0.75 & 1.50 & 4.88 & 0.94 & -1.04 & 1.72 & 4.61 & 0.94\\
        
        \multirow{-3}{*}{S4} & ATE & 1.76 & 0.05 & 4.59 & 8.17 & 0.94 & 0.19 & 2.28 & 5.59 & 0.94 & 0.06 & 1.18 & 4.14 & 0.94 & 0.00 & 0.42 & 3.61 & 0.99\\
        \cmidrule{1-19}
        & NIE & 1.64 & 0.40 & 2.58 & 5.47 & 0.91 & 0.42 & 1.05 & 4.10 & 0.95 & 0.52 & 0.87 & 4.26 & 0.96 & 0.64 & 1.10 & 3.39 & 0.91\\
        
        & NDE & 2.52 & -0.20 & 5.67 & 10.14 & 0.96 & -0.35 & 2.48 & 6.80 & 0.95 & -0.54 & 1.54 & 5.41 & 0.96 & -0.58 & 1.14 & 4.60 & 0.97\\
        
        \multirow{-3}{*}{S5} & ATE & 4.17 & 0.20 & 5.99 & 9.30 & 0.94 & 0.22 & 2.93 & 6.32 & 0.93 & 0.02 & 1.54 & 4.74 & 0.94 & 0.03 & 0.54 & 4.23 & 0.99\\
        \cmidrule{1-19}
        & NIE & 0.18 & 0.15 & 0.04 & 0.63 & 0.90 & 0.15 & 0.03 & 0.49 & 0.82 & 0.16 & 0.03 & 0.39 & 0.65 & 0.14 & 0.02 & 0.32 & 0.61\\
        
        & NDE & 2.39 & -0.22 & 0.07 & 0.84 & 0.89 & -0.20 & 0.05 & 0.54 & 0.79 & -0.18 & 0.03 & 0.36 & 0.52 & -0.16 & 0.03 & 0.23 & 0.11\\
        
        \multirow{-3}{*}{S6} & ATE & 2.57 & -0.05 & 0.03 & 0.81 & 0.97 & -0.05 & 0.02 & 0.57 & 0.96 & -0.02 & 0.01 & 0.44 & 0.99 & -0.01 & 0.00 & 0.35 & 1.00\\
        \cmidrule{1-19}
        & NIE & 1.19 & 0.71 & 1.43 & 3.42 & 0.84 & 0.70 & 0.87 & 2.36 & 0.76 & 0.71 & 0.69 & 1.72 & 0.60 & 0.73 & 0.62 & 1.20 & 0.26\\
        
        & NDE & 0.95 & -0.71 & 2.40 & 6.29 & 0.93 & -0.70 & 1.09 & 3.00 & 0.86 & -0.73 & 0.90 & 2.27 & 0.77 & -0.76 & 0.72 & 1.51 & 0.52\\
        
        \multirow{-3}{*}{S7} & ATE & 2.14 & 0.08 & 2.30 & 5.92 & 0.95 & 0.03 & 1.12 & 3.95 & 0.94 & -0.02 & 0.58 & 2.93 & 0.94 & 0.00 & 0.23 & 2.00 & 0.95\\
        \cmidrule{1-19}
        & NIE & 1.89 & 0.66 & 1.89 & 4.19 & 0.86 & 0.67 & 1.02 & 2.88 & 0.83 & 0.68 & 0.75 & 2.14 & 0.73 & 0.71 & 0.62 & 1.53 & 0.53\\
        
        & NDE & 1.74 & -0.67 & 2.60 & 4.81 & 0.92 & -0.72 & 1.24 & 3.26 & 0.88 & -0.73 & 0.97 & 2.44 & 0.80 & -0.76 & 0.74 & 1.64 & 0.60\\
        
        \multirow{-3}{*}{S8} & ATE & 3.63 & 0.10 & 3.03 & 6.38 & 0.93 & -0.02 & 1.51 & 4.50 & 0.93 & -0.03 & 0.78 & 3.42 & 0.93 & -0.01 & 0.32 & 2.35 & 0.95\\
        \cmidrule{1-19}
        & NIE & 0.30 & 0.08 & 0.04 & 0.68 & 0.93 & 0.07 & 0.02 & 0.49 & 0.92 & 0.03 & 0.01 & 0.38 & 0.95 & 0.02 & 0.00 & 0.29 & 0.98\\
        
        & NDE & 2.51 & -0.13 & 0.03 & 0.48 & 0.82 & -0.13 & 0.02 & 0.32 & 0.65 & -0.10 & 0.01 & 0.21 & 0.50 & -0.10 & 0.01 & 0.13 & 0.13\\
        
        \multirow{-3}{*}{S9} & ATE & 2.81 & -0.05 & 0.04 & 0.73 & 0.94 & -0.06 & 0.02 & 0.52 & 0.94 & -0.08 & 0.01 & 0.39 & 0.91 & -0.08 & 0.01 & 0.29 & 0.85\\
        \cmidrule{1-19}
        & NIE & 1.17 & 0.24 & 1.37 & 3.83 & 0.90 & 0.19 & 0.42 & 2.40 & 0.95 & 0.21 & 0.26 & 1.76 & 0.93 & 0.22 & 0.13 & 1.21 & 0.90\\
        
        & NDE & 0.92 & -0.21 & 2.82 & 7.15 & 0.96 & -0.18 & 0.82 & 3.37 & 0.94 & -0.20 & 0.51 & 2.51 & 0.94 & -0.22 & 0.24 & 1.72 & 0.91\\
        
        \multirow{-3}{*}{S10} & ATE & 2.10 & 0.09 & 2.36 & 6.19 & 0.95 & 0.03 & 1.02 & 3.79 & 0.93 & 0.01 & 0.57 & 2.82 & 0.94 & 0.03 & 0.22 & 1.92 & 0.97\\
        \cmidrule{1-19}
        & NIE & 1.86 & 0.18 & 1.83 & 4.26 & 0.90 & 0.15 & 0.59 & 2.91 & 0.94 & 0.18 & 0.37 & 2.16 & 0.93 & 0.23 & 0.19 & 1.49 & 0.91\\
        
        & NDE & 1.77 & -0.22 & 3.11 & 7.35 & 0.94 & -0.20 & 0.95 & 3.54 & 0.93 & -0.22 & 0.62 & 2.71 & 0.93 & -0.22 & 0.28 & 1.85 & 0.92\\
        
        \multirow{-3}{*}{S11} & ATE & 3.63 & 0.04 & 2.97 & 6.61 & 0.94 & -0.01 & 1.36 & 4.28 & 0.94 & 0.02 & 0.77 & 3.29 & 0.93 & 0.03 & 0.29 & 2.27 & 0.97\\
        \cmidrule{1-19}
        & NIE & 0.26 & 0.03 & 0.02 & 0.50 & 0.93 & 0.00 & 0.01 & 0.36 & 0.94 & -0.01 & 0.00 & 0.28 & 0.95 & -0.02 & 0.00 & 0.23 & 0.99\\
        
        & NDE & 2.50 & -0.08 & 0.01 & 0.34 & 0.84 & -0.06 & 0.01 & 0.21 & 0.80 & -0.05 & 0.00 & 0.14 & 0.72 & -0.04 & 0.00 & 0.08 & 0.51\\
        
        \multirow{-3}{*}{S12} & ATE & 2.76 & -0.05 & 0.02 & 0.56 & 0.92 & -0.06 & 0.01 & 0.39 & 0.91 & -0.06 & 0.01 & 0.30 & 0.90 & -0.06 & 0.01 & 0.23 & 0.90\\
        \bottomrule
    \end{tabular}
\end{table} 
    \begin{table}[htbp]
    \fontsize{6}{11}\selectfont
    \centering
    \caption{\label{EDPM:tab:sim:para} Simulation results for NIE, NDE, and ATE with a post-treatment variable under the parametric Bayesian model. over 500 replications. The columns correspond to bias, MSE, CIl (length of 95\% credible interval), and CP (the empirical coverage probability).}
    \begin{tabular}[htbp]{llr|rrrr|rrrr|rrrr|rrrr} 
        \toprule
        \multicolumn{3}{c}{ } & \multicolumn{4}{c}{250} & \multicolumn{4}{c}{500} & \multicolumn{4}{c}{1000} & \multicolumn{4}{c}{2500} \\
        \cmidrule(l{3pt}r{3pt}){4-7} \cmidrule(l{3pt}r{3pt}){8-11} \cmidrule(l{3pt}r{3pt}){12-15} \cmidrule(l{3pt}r{3pt}){16-19}
        \midrule
        Scn & CE & Truth & Bias & MSE & CIl & CP & Bias & MSE & CIl & CP & Bias & MSE & CIl & CP & Bias & MSE & CIl & CP \\
        & NIE & 1.19 & -0.02 & 1.93 & 4.63 & 0.89 & -0.08 & 0.85 & 3.09 & 0.89 & -0.02 & 0.42 & 2.17 & 0.90 & -0.03 & 0.17 & 1.37 & 0.92\\
        
        & NDE & 0.95 & -0.05 & 5.56 & 9.17 & 0.95 & 0.10 & 2.60 & 6.30 & 0.95 & -0.04 & 1.37 & 4.49 & 0.94 & -0.03 & 0.48 & 2.88 & 0.96\\
        
        \multirow{-3}{*}{S1} & ATE & 2.14 & 0.06 & 4.55 & 8.17 & 0.94 & 0.03 & 2.25 & 5.66 & 0.94 & -0.03 & 1.14 & 4.01 & 0.95 & 0.02 & 0.40 & 2.59 & 0.96\\
        \cmidrule{1-19}
        & NIE & 1.89 & -0.14 & 3.73 & 5.84 & 0.85 & -0.01 & 1.63 & 3.93 & 0.86 & -0.05 & 0.75 & 2.76 & 0.89 & -0.03 & 0.35 & 1.76 & 0.86\\
        
        & NDE & 2.56 & -0.05 & 7.11 & 10.69 & 0.95 & 0.08 & 3.25 & 7.37 & 0.95 & -0.08 & 1.71 & 5.23 & 0.95 & 0.00 & 0.59 & 3.35 & 0.98\\
        
        \multirow{-3}{*}{S2} & ATE & 4.45 & 0.10 & 5.98 & 9.37 & 0.93 & 0.08 & 2.99 & 6.51 & 0.93 & -0.03 & 1.47 & 4.59 & 0.94 & 0.01 & 0.52 & 2.97 & 0.96\\
        \cmidrule{1-19}
        & NIE & 0.30 & -0.14 & 0.05 & 0.42 & 0.66 & -0.15 & 0.04 & 0.29 & 0.51 & -0.14 & 0.03 & 0.20 & 0.39 & -0.13 & 0.02 & 0.13 & 0.16\\
        
        & NDE & 2.53 & 0.11 & 0.08 & 1.04 & 0.92 & 0.11 & 0.05 & 0.74 & 0.88 & 0.13 & 0.04 & 0.54 & 0.86 & 0.13 & 0.03 & 0.37 & 0.73\\
        
        \multirow{-3}{*}{S3} & ATE & 2.83 & -0.01 & 0.07 & 0.99 & 0.94 & -0.02 & 0.04 & 0.71 & 0.92 & -0.01 & 0.02 & 0.52 & 0.94 & 0.00 & 0.01 & 0.36 & 0.97\\
        \cmidrule{1-19}
        & NIE & 1.04 & -0.04 & 1.50 & 4.04 & 0.89 & -0.06 & 0.65 & 2.73 & 0.89 & -0.01 & 0.36 & 1.91 & 0.88 & 0.00 & 0.14 & 1.22 & 0.90\\
        
        & NDE & 0.72 & 0.06 & 5.32 & 9.09 & 0.94 & 0.13 & 2.51 & 6.30 & 0.96 & 0.00 & 1.28 & 4.41 & 0.95 & -0.03 & 0.45 & 2.82 & 0.96\\
        
        \multirow{-3}{*}{S4} & ATE & 1.76 & 0.00 & 4.69 & 8.30 & 0.93 & 0.14 & 2.32 & 5.76 & 0.93 & 0.01 & 1.24 & 4.05 & 0.93 & 0.01 & 0.43 & 2.58 & 0.95\\
        \cmidrule{1-19}
        & NIE & 1.64 & -0.06 & 3.08 & 5.18 & 0.83 & -0.02 & 1.29 & 3.52 & 0.86 & -0.06 & 0.72 & 2.46 & 0.84 & 0.01 & 0.26 & 1.57 & 0.87\\
        
        & NDE & 2.52 & 0.21 & 6.91 & 10.55 & 0.95 & 0.07 & 3.10 & 7.30 & 0.96 & 0.03 & 1.60 & 5.10 & 0.96 & 0.00 & 0.58 & 3.27 & 0.96\\
        
        \multirow{-3}{*}{S5} & ATE & 4.17 & 0.16 & 5.92 & 9.40 & 0.94 & 0.16 & 2.96 & 6.52 & 0.94 & -0.01 & 1.55 & 4.62 & 0.94 & 0.01 & 0.53 & 2.94 & 0.96\\
        \cmidrule{1-19}
        & NIE & 0.18 & -0.02 & 0.02 & 0.38 & 0.78 & -0.02 & 0.01 & 0.26 & 0.76 & -0.01 & 0.01 & 0.18 & 0.77 & -0.01 & 0.00 & 0.12 & 0.79\\
        
        & NDE & 2.39 & -0.01 & 0.06 & 0.96 & 0.95 & 0.00 & 0.04 & 0.69 & 0.92 & 0.00 & 0.02 & 0.50 & 0.93 & 0.01 & 0.01 & 0.34 & 0.96\\
        
        \multirow{-3}{*}{S6} & ATE & 2.57 & -0.01 & 0.06 & 0.91 & 0.94 & -0.01 & 0.03 & 0.65 & 0.90 & 0.00 & 0.02 & 0.48 & 0.96 & 0.00 & 0.01 & 0.32 & 0.97\\
        \cmidrule{1-19}
        & NIE & 1.19 & -0.02 & 1.66 & 4.19 & 0.86 & -0.04 & 0.83 & 2.82 & 0.88 & -0.01 & 0.46 & 1.95 & 0.83 & -0.01 & 0.19 & 1.25 & 0.84\\
        
        & NDE & 0.95 & 0.06 & 3.42 & 7.58 & 0.96 & 0.05 & 1.54 & 5.19 & 0.95 & 0.00 & 0.81 & 3.63 & 0.95 & 0.02 & 0.29 & 2.31 & 0.95\\
        
        \multirow{-3}{*}{S7} & ATE & 2.14 & 0.08 & 2.74 & 6.53 & 0.95 & -0.03 & 1.37 & 4.45 & 0.93 & -0.01 & 0.69 & 3.15 & 0.93 & 0.01 & 0.26 & 2.00 & 0.95\\
        \cmidrule{1-19}
        & NIE & 1.89 & -0.05 & 2.67 & 4.91 & 0.85 & -0.07 & 1.36 & 3.32 & 0.84 & -0.02 & 0.75 & 2.32 & 0.81 & 0.00 & 0.31 & 1.49 & 0.82\\
        
        & NDE & 1.74 & 0.05 & 4.40 & 8.76 & 0.95 & 0.05 & 2.04 & 5.99 & 0.96 & 0.05 & 1.03 & 4.19 & 0.97 & 0.02 & 0.37 & 2.66 & 0.96\\
        
        \multirow{-3}{*}{S8} & ATE & 3.63 & 0.08 & 3.67 & 7.52 & 0.94 & 0.00 & 1.85 & 5.17 & 0.95 & -0.02 & 0.93 & 3.63 & 0.92 & 0.01 & 0.34 & 2.32 & 0.95\\
        \cmidrule{1-19}
        & NIE & 0.30 & -0.02 & 0.02 & 0.49 & 0.91 & -0.02 & 0.01 & 0.35 & 0.93 & -0.02 & 0.01 & 0.27 & 0.94 & -0.02 & 0.00 & 0.20 & 0.97\\
        
        & NDE & 2.51 & -0.02 & 0.01 & 0.51 & 0.97 & -0.03 & 0.01 & 0.38 & 0.97 & -0.03 & 0.00 & 0.30 & 0.98 & -0.03 & 0.00 & 0.25 & 1.00\\
        
        \multirow{-3}{*}{S9} & ATE & 2.81 & -0.05 & 0.04 & 0.73 & 0.95 & -0.05 & 0.02 & 0.53 & 0.94 & -0.04 & 0.01 & 0.40 & 0.95 & -0.04 & 0.01 & 0.30 & 0.96\\
        \cmidrule{1-19}
        & NIE & 1.17 & -0.05 & 1.63 & 4.14 & 0.87 & -0.09 & 0.85 & 2.82 & 0.87 & -0.02 & 0.45 & 1.96 & 0.84 & 0.03 & 0.20 & 1.24 & 0.83\\
        
        & NDE & 0.92 & 0.09 & 3.44 & 7.48 & 0.96 & 0.03 & 1.50 & 5.13 & 0.96 & 0.00 & 0.82 & 3.57 & 0.95 & -0.01 & 0.31 & 2.26 & 0.95\\
        
        \multirow{-3}{*}{S10} & ATE & 2.10 & 0.16 & 2.66 & 6.37 & 0.95 & 0.01 & 1.28 & 4.36 & 0.94 & 0.01 & 0.70 & 3.03 & 0.93 & 0.03 & 0.24 & 1.93 & 0.94\\
        \cmidrule{1-19}
        & NIE & 1.86 & -0.03 & 2.64 & 4.85 & 0.85 & -0.10 & 1.39 & 3.28 & 0.83 & -0.03 & 0.73 & 2.28 & 0.81 & 0.03 & 0.33 & 1.47 & 0.80\\
        
        & NDE & 1.77 & 0.16 & 4.42 & 8.59 & 0.96 & 0.03 & 1.97 & 5.89 & 0.96 & 0.02 & 1.05 & 4.12 & 0.95 & -0.02 & 0.41 & 2.59 & 0.95\\
        
        \multirow{-3}{*}{S11} & ATE & 3.63 & 0.14 & 3.51 & 7.32 & 0.94 & 0.00 & 1.71 & 5.03 & 0.94 & 0.04 & 0.93 & 3.50 & 0.93 & 0.03 & 0.32 & 2.22 & 0.94\\
        \cmidrule{1-19}
        & NIE & 0.26 & 0.02 & 0.02 & 0.47 & 0.93 & 0.02 & 0.01 & 0.34 & 0.91 & 0.02 & 0.00 & 0.26 & 0.94 & 0.03 & 0.00 & 0.20 & 0.96\\
        
        & NDE & 2.50 & -0.07 & 0.01 & 0.39 & 0.90 & -0.08 & 0.01 & 0.29 & 0.84 & -0.08 & 0.01 & 0.23 & 0.82 & -0.08 & 0.01 & 0.19 & 0.73\\
        
        \multirow{-3}{*}{S12} & ATE & 2.76 & -0.05 & 0.02 & 0.57 & 0.94 & -0.05 & 0.01 & 0.42 & 0.93 & -0.05 & 0.01 & 0.31 & 0.93 & -0.04 & 0.00 & 0.23 & 0.94\\
        \bottomrule
    \end{tabular}
\end{table}

    \begin{table}[htbp]
    \centering
    \caption{\label{EDPM:tab:avgmain} Posterior means (95\% credible intervals) of NIE, NDE, and ATE under the proposed BNP model. $\rho$ is a sensitivity parameter for a post-treatment confounder.}
    \begin{tabular}[htbp]{c|rrr|rrr|rrr}
        \toprule
        \multicolumn{1}{c}{ } & \multicolumn{3}{c}{$\rho \sim \munif{-1,0}$} & \multicolumn{3}{c}{$\rho \sim \munif{-1,1}$} & \multicolumn{3}{c}{$\rho \sim \mtri{0,1,1}$} \\
        \cmidrule(l{3pt}r{3pt}){1-1} \cmidrule(l{3pt}r{3pt}){2-4} \cmidrule(l{3pt}r{3pt}){5-7} \cmidrule(l{3pt}r{3pt}){8-10}
        CE & Est. & \multicolumn{2}{c}{95\% CI} & Est. & \multicolumn{2}{c}{95\% CI} & Est. & \multicolumn{2}{c}{95\% CI} \\
        \cmidrule(l{3pt}r{3pt}){1-1} \cmidrule(l{3pt}r{3pt}){2-4} \cmidrule(l{3pt}r{3pt}){5-7} \cmidrule(l{3pt}r{3pt}){8-10}
        NIE & -0.82 & -1.44 & -0.26 & -0.57 & -1.24 & 0.02 & -0.29 & -0.76 & 0.15 \\
        NDE & -2.53 & -3.98 & -1.08 & -2.74 & -4.28 & -1.23 & -3.02 & -4.46 & -1.64 \\
        ATE & -3.34 & -4.75 & -2.01 & -3.31 & -4.76 & -1.94 & -3.31 & -4.77 & -1.91 \\
        \midrule
        
        \multicolumn{1}{c}{} & \multicolumn{3}{c}{$\rho = - 0.8$} & \multicolumn{3}{c}{$\rho = 0$} & \multicolumn{3}{c}{$\rho = 0.5$} \\
        \cmidrule(l{3pt}r{3pt}){1-1} \cmidrule(l{3pt}r{3pt}){2-4} \cmidrule(l{3pt}r{3pt}){5-7} \cmidrule(l{3pt}r{3pt}){8-10}
        CE & Est. & \multicolumn{2}{c}{95\% CI} & Est. & \multicolumn{2}{c}{95\% CI} & Est. & \multicolumn{2}{c}{95\% CI} \\
        \cmidrule(l{3pt}r{3pt}){1-1} \cmidrule(l{3pt}r{3pt}){2-4} \cmidrule(l{3pt}r{3pt}){5-7} \cmidrule(l{3pt}r{3pt}){8-10}
        NIE & -0.93 & -1.44 & -0.38 & -0.59 & -1.03 & -0.16 & -0.38 & -0.79 & 0.07 \\
        NDE & -2.38 & -3.87 & -0.87 & -2.73 & -4.19 & -1.36 & -2.97 & -4.41 & -1.58 \\
        ATE & -3.30 & -4.79 & -1.94 & -3.32 & -4.74 & -1.95 & -3.34 & -4.77 & -1.92 \\
        \bottomrule
    \end{tabular}
\end{table}
    \begin{table}[htbp]
    \centering
    \caption{\label{EDPM:tab:condmain} Posterior means (95\% credible intervals) of NIE, NDE, and ATE conditional on race (B=Black, W=White) under the proposed BNP model. Other covariates are integrated out. $\rho$ is a sensitivity parameter for a post-treatment confounder.}
    \begin{tabular}[htbp]{cc|rrr|rrr|rrr}
        \toprule
        \multicolumn{1}{c}{ } & \multicolumn{1}{c}{ } & \multicolumn{3}{c}{$\rho \sim \munif{-1,0}$} & \multicolumn{3}{c}{$\rho \sim \munif{-1,1}$} & \multicolumn{3}{c}{$\rho \sim \mtri{0,1,1}$} \\
        \cmidrule(l{3pt}r{3pt}){1-1} \cmidrule(l{3pt}r{3pt}){2-2} \cmidrule(l{3pt}r{3pt}){3-5} \cmidrule(l{3pt}r{3pt}){6-8} \cmidrule(l{3pt}r{3pt}){9-11}
        Race & CE & Est. & \multicolumn{2}{c}{95\% CI} & Est. & \multicolumn{2}{c}{95\% CI} & Est. & \multicolumn{2}{c}{95\% CI} \\
        
        \cmidrule(l{3pt}r{3pt}){1-1} \cmidrule(l{3pt}r{3pt}){2-2} \cmidrule(l{3pt}r{3pt}){3-5} \cmidrule(l{3pt}r{3pt}){6-8} \cmidrule(l{3pt}r{3pt}){9-11}
         & NIE & -0.11 & -0.56 & 0.35 & -0.10 & -0.57 & 0.31 & -0.08 & -0.56 & 0.35\\
         & NDE & -2.67 & -4.22 & -1.11 & -2.66 & -4.25 & -1.10 & -2.67 & -4.17 & -1.14\\
        \multirow{-3}{*}{B} & ATE & -2.78 & -4.43 & -1.13 & -2.76 & -4.31 & -1.12 & -2.75 & -4.36 & -1.09\\
        
        \cmidrule(l{3pt}r{3pt}){1-1} \cmidrule(l{3pt}r{3pt}){2-2} \cmidrule(l{3pt}r{3pt}){3-5} \cmidrule(l{3pt}r{3pt}){6-8} \cmidrule(l{3pt}r{3pt}){9-11}
         & NIE & -0.42 & -0.93 & 0.04 & -0.41 & -0.94 & 0.13 & -0.42 & -0.95 & 0.05\\
         & NDE & -3.02 & -4.47 & -1.55 & -3.04 & -4.64 & -1.62 & -3.03 & -4.62 & -1.56\\
        \multirow{-3}{*}{W} & ATE & -3.44 & -4.93 & -1.95 & -3.45 & -4.97 & -1.95 & -3.44 & -4.96 & -1.98\\
        
        \midrule
        
        \multicolumn{1}{c}{} & \multicolumn{1}{c}{} & \multicolumn{3}{c}{$\rho = - 0.8$} & \multicolumn{3}{c}{$\rho = 0$} & \multicolumn{3}{c}{$\rho = 0.5$} \\
        \cmidrule(l{3pt}r{3pt}){1-1} \cmidrule(l{3pt}r{3pt}){2-2} \cmidrule(l{3pt}r{3pt}){3-5} \cmidrule(l{3pt}r{3pt}){6-8} \cmidrule(l{3pt}r{3pt}){9-11}
        
         & NIE & -0.11 & -0.60 & 0.34 & -0.10 & -0.61 & 0.31 & -0.14 & -0.60 & 0.35\\
         & NDE & -2.70 & -4.25 & -1.13 & -2.71 & -4.21 & -1.13 & -2.64 & -4.20 & -1.03\\
        \multirow{-3}{*}{B} & ATE & -2.80 & -4.48 & -1.18 & -2.81 & -4.37 & -1.17 & -2.78 & -4.33 & -1.14\\
        
        \cmidrule(l{3pt}r{3pt}){1-1} \cmidrule(l{3pt}r{3pt}){2-2} \cmidrule(l{3pt}r{3pt}){3-5} \cmidrule(l{3pt}r{3pt}){6-8} \cmidrule(l{3pt}r{3pt}){9-11}
         & NIE & -0.41 & -0.99 & 0.08 & -0.41 & -0.92 & 0.09 & -0.42 & -0.96 & 0.07\\
         & NDE & -3.05 & -4.61 & -1.59 & -3.03 & -4.57 & -1.60 & -3.03 & -4.62 & -1.55\\
        \multirow{-3}{*}{W} & ATE & -3.46 & -5.03 & -2.03 & -3.44 & -4.96 & -2.01 & -3.44 & -4.98 & -1.97\\
        \bottomrule
    \end{tabular}
\end{table} 
    
\end{document}